\documentclass[twocolumn]{aa}  

\usepackage{mathrsfs}
\usepackage{amsmath}

\usepackage{graphicx}
\usepackage{txfonts}
\usepackage{color}
\usepackage{xcolor}
\usepackage{multicol}
\usepackage{makecell}
\usepackage{subcaption}
\usepackage{pdflscape}
\usepackage{longtable}
\usepackage{booktabs}
\usepackage{caption}
\usepackage[skip=2pt plus1pt, indent=15pt]{parskip}
\usepackage[flushleft]{threeparttable}
\usepackage{tikz,caption}
\usepackage{hyperref}
\usetikzlibrary{calc}
\DeclareCaptionFormat{overlay}{\gdef\capoverlay{#1#2#3\par}}
\DeclareCaptionLabelSeparator{frcolon}{ $\,:$ }
\DeclareCaptionStyle{overlay}{format=overlay}
\usepackage{csvsimple}
\usepackage{supertabular}

\hypersetup{
    colorlinks=true,
    linkcolor=blue,
    citecolor=blue,     
    urlcolor=blue,
    linkbordercolor={red}
}

\begin{document}

   \title{Low cosmic-ray ionisation at parsec scales in G035.39-00.33}

   %\subtitle{}
   % OR Density dependence of the fiber widths

   %\titlerunning{}

   \author{A. Socci\inst{1}
          \and
          A. Barnes\inst{2}
          \and
          J. Pineda\inst{3}
          \and
          O. Sipil\"{a}\inst{3}
          \and
          C. Gieser\inst{4}
          \and
          P. Caselli\inst{3}
          \and
          A. Hacar\inst{1}
          \and
          G. Cosentino\inst{5}
          \and
          C. Y. Law\inst{6}
          \and
          M. T. Owen\inst{7}
          \and
          P. Gorai\inst{8,9}
          \and
          F. Fontani\inst{3,6,10}
          \and
          I. Jim\'{e}nez-Serra\inst{11}
          \and
          J. C. Tan\inst{12,13}
          }

   \institute{1 - Institute for Astronomy (IfA), University of Vienna,
              T\"{u}rkenschanzstrasse 17, A-1180 Vienna\\
              \email{andrea.socci@univie.ac.at} \\
              2 - European Southern Observatory (ESO), Karl-Schwarzschild-Stra{\ss}e 2, 85748 Garching bei M\"{u}nchen, Germany\\
              3 - Max Planck Institute for Extraterrestrial Physics (MPE), Gie{\ss}enbachstraße 1, 85748 Garching bei M\"{u}nchen, Germany\\
              4 - Max Planck Institute for Astronomy, K\"{o}nigstuhl 17, 69117 Heidelberg, Germany\\
              5 - Institut de Radioastronomie Millim\'{e}trique (IRAM), 300 Rue de la Piscine, 38400 Saint-Martin-d’H\'{e}res, France\\
              6 - INAF Osservatorio Astrofisico di Arcetri, Largo Enrico Fermi, 5, 50125 Firenze FI, Italy\\
              7 - Physikalisches Institut, Universit\"{a}t zu K\"{o}ln, Z\"{u}lpicher Stra{\ss}e 77, 50937 K\"{o}ln, Germany\\
              8 - Rosseland Centre for Solar Physics, University of Oslo, PO Box 1029 Blindern, 0315 Oslo, Norway\\
              9 - Institute of Theoretical Astrophysics, University of Oslo, PO Box 1029 Blindern, 0315 Oslo, Norway\\
              10 - Laboratory for the study of the Universe and eXtreme phenomena (LUX), Observatoire de Paris, PSL Research University, CNRS, Sorbonne Universit\'{e}, 92190 Meudon, France\\
              11 - Centro de Astrobiolog\'{i}a (CSIC/INTA), Ctra. de Torrej\'{o}na Ajalvir km 4, Madrid, Spain\\
              12 - Department of Physics \& Astronomy, Chalmers University of Technology, 412 96 Gothenburg, Sweden\\
              13 - Department of Astronomy, University of Virginia, 530 McCormick Road, Charlottesville, VA 22904-4325, USA
             }

   \date{Received; -- accepted; --}

% \abstract{}{}{}{}{} 
% 5 {} token are mandatory
 
  \abstract
  % Context {} leave it empty if necessary  
   {Cosmic rays (CRs) play a key role in the interstellar medium (ISM) by regulating the chemical evolution of the gas and its coupling to the magnetic field in the densest and coldest regions of molecular clouds. However, the CR ionisation rate of H$_2$ ($\zeta_2$) is one of the most debated parameters characterising molecular clouds due to the uncertainties in its estimation.
   }
  % aims heading (mandatory)
   {We aim to derive a new and reliable indirect analytical method to probe the electron fraction, $x(e)$, and $\zeta_2$ of the gas across multiple density regimes. We further apply this novel framework to a parsec-scale filament to homogeneously map a significant sample of fields with different physical conditions and test their impact on $x(e)$ and $\zeta_2$.
   }
  % methods heading (mandatory)
   {Recent estimates of $x(e)$ and $\zeta_2$ suffer from limitations, mostly driven by observational restrictions. We thus developed a new analytical framework based on the chemistry of N$_2$H$^+$, N$_2$D$^+$ and DCO$^+$. We mapped the emission from their ground transition towards the parsec-scale filament of the infrared dark cloud (IRDC) G035.39-00.33 with new observations from the NOrthern Extended Array (NOEMA) at a resolution of $3''$ (or $\sim9000$~au at the cloud distance). By combining this novel survey with ancillary observations of C$^{18}$O ($1-0$), we determined the CO depletion factor, the N$_2$H$^+$ deuterium fraction and, ultimately, $x(e)$ and $\zeta_2$ in G035.39-00.33.
   }
  % results heading (mandatory)
   {The CO depletion is significant and widespread in G035.39-00.33. Its depletion factor, $f_\mathrm{D}$, shows a positive correlation with column and number densities of H$_2$ in the cloud, with the highest intensities in both N$_2$H$^+$ and N$_2$D$^+$, and with the sites displaying an enhanced deuterium fraction ($R_\mathrm{D}$). The electron fraction varies by two orders of magnitude within $\sim10^{-9}-10^{-7}$ and shows a functional dependence on the number density of H$_2$ in the region. This correlation is consistent with that determined in low-mass cores with similar degrees of CO depletion, but now extended to parsec scales, and with the predictions from chemical models. $\zeta_2$ varies by three orders of magnitude within $\sim10^{-18}-10^{-15}$~s$^{-1}$, and most fields show values of $\sim2.3\times10^{-18}$~s$^{-1}$ on average. $\zeta_2$ shows a functional dependence on $N(\mathrm{H_2})$, as predicted by theoretical models, but its scatter and its values, on average lower than the typical $\zeta_2$ for the ISM, suggest the presence of local enhancement and overall attenuation of the CR flux taking place in G035.39-00.33.
   }
   %conclusions
   {$f_\mathrm{D}$ and $R_\mathrm{D}$ show values consistent with theoretical predictions and previous independent studies. On the other hand, $x(e)$ and $\zeta_2$ show values at the lower end compared to independent estimates in filaments and cores. The values of $\zeta_2$ are consistent with those reported for other IRDCs and giant filaments, but lower compared to theoretical predictions for the column density regime sampled by G035.39-00.33. CR attenuation can come from the change in magnetic field strength and morphology previously reported in G035.39-00.33, which explains the reduced values of $x(e)$ and $\zeta_2$. Fields with ionisation rates of $\zeta_2\gtrsim10^{-16}$~s$^{-1}$ may be affected by the local acceleration of CRs or by the magnetic pockets formed from the uneven magnetic field instead.
   }

   \keywords{Interstellar Medium -- Massive star formation -- Astrochemistry -- Instrumentation: interferometers -- ISM: cosmic rays -- ISM: molecules -- ISM: structure
               }

   \maketitle
%
%-------------------------------------------------------------------
\renewcommand{\ttdefault}{pcr}
\captionsetup{labelfont=bf}

\section{Introduction}\label{sec:introduction}

%the relevance of cosmic rays
Cosmic rays (CRs) are charged accelerated particles that play a key role in the evolution of the interstellar medium (ISM) \citep{padovani2020}. At densities $n\gtrsim10^3$~cm$^{-3}$, low-energy CRs ($E<10$~GeV) mostly interact with H$_2$, causing its ionisation at a rate $\zeta_2$. This ionisation rate, in turn, triggers many processes within molecular clouds. First, H$_2^+$ rapidly forms H$_3^+$ \citep{herbst1973}, initiating ion-neutral chemistry in dense clouds and allowing the formation of fundamental ions and complex molecules \citep{herbst2009}. Recombination of these ions with electrons and neutrals is the main source of heating in molecular clouds and filaments \citep{glassgold2012}, thus setting the thermal pressure that confines the dense cores embedded within them \citep{goldsmith2001}. The balance between ions and neutrals, usually expressed via the electron fraction $x(e)=n(e)/n$ (with $n(e)$ and $n$ number densities of electrons and total gas, respectively), influences the coupling between the gas and magnetic field \citep{mckee1989} with the resulting magnetic pressure potentially slowing the gravitational collapse of individual cores \citep{shu1987}. Ultimately, CRs play a pivotal role in shaping the initial conditions for star formation through their ionisation rate $\zeta_2$.

%the elusiveness of the CR ionisation rate
Measuring $\zeta_2$ is however complex. Its most direct probe is the abundance of H$_3^+$. This species has several vibrational transitions in the infrared (IR) \citep{oka1980}, where detections have been reported since the 1990s \citep[see][]{oka2019}. As a homonuclear and symmetric molecule, H$_3^+$ lacks rotational transitions in the millimetre regime (mm), where emission from the cold gas of molecular clouds peaks instead. The elusiveness of H$_3^+$ in these regions has so far prevented direct and accurate measurements of $\zeta_2$ towards the sites of star formation.

%the advent of proxies and their variety
Proxies were therefore used over the years to determine $\zeta_2$ in the ISM. Early studies using radicals \citep{black1977}, neutrals \citep{vandishoeck1986}, and ions \citep{vandertak2000} reported a homogeneous value of $\sim10^{-17}$~s$^{-1}$ across the ISM, consistent with the theoretical prediction based on the CR spectrum \citep{spitzer1968,webber1998}. During the past 30 years, studies in dense cores within low-mass clouds \citep{caselli1998}, intermediate-mass envelopes \citep[e.g.][]{ceccarelli2014,fontani2017}, and high-mass regions \citep[e.g.][]{luo2024} reported ionisation rates varying by up to four orders of magnitude ($\zeta_2\sim10^{-18}-10^{-14}$~s$^{-1}$). Inhomogeneous sampling of $\zeta_2$ from scales of $\sim0.1$~pc \citep{redaelli2021} up to $\sim1$~pc \citep[][]{socci2024a} with tracers sensitive to selected environments \citep[evolved vs pre-stellar;][respectively]{luo2024,bovino2020} made the link of $\zeta_2$ to the physical (e.g. density) and environmental (e.g. feedback) properties of the region more complex.

%two methods to estimate zeta in maps
Recently, two methods proved reliable in mapping $\zeta_2$ across scales. The first method, proposed by \citet{bovino2020}, is based on the observations of H$_2$D$^+$, the product of the first deuteration of H$_3^+$ \citep{watson1974}. Together with observations of DCO$^+$, HCO$^+$, and CO, this method allowed \citet{sabatini2023} to probe $\zeta_2$ in two massive clumps at scales of $\sim0.005-0.1$~pc. The second method, proposed by \citet{caselli1998}, uses observations of DCO$^+$, HCO$^+$, and CO, plus assumptions on the abundances of species that cannot be observed in the dense gas (e.g. O, HD). This method allowed \citet{pineda2024a} to probe $\zeta_2$ in a low-mass proto-cluster on scales of $\sim0.01-0.5$~pc. Both studies reported a variation of $\zeta_2$ in the regions related either to a change in the column density of H$_2$, $N(\mathrm{H_2})$, or to the production of CRs from local sources (i.e. protostars), in agreement with predictions from theoretical models \citep[][respectively]{padovani2009,padovani2016}.

%the limitation of these two methods
Despite connecting the large dynamic range in $\zeta_2$ to the properties of the region in a self-consistent way, these methods still suffer from limitations, mostly related to observations \citep[see][for a full discussion on the two methods]{redaelli2024}. The method of \citet{bovino2020} depends on the abundance of H$_2$D$^+$ whose ortho spin state is the only one with a transition in the millimetre regime (at $\sim372$~GHz). Comprising only a fraction of its total (ortho and para) abundance, ortho-H$_2$D$^+$ is rarely extended in clouds and its detection becomes challenging on scales $\gtrsim0.2$~pc. The method of \citet{caselli1998} depends on the number density of H$_2$, $n(\mathrm{H_2})$, which is complex to measure in clouds. A poorly estimated $n(\mathrm{H_2})$ could cause $\zeta_2$ to increase for increasing density, in contradiction with the theory of CR propagation in a progressively denser medium. Finally, both methods require an estimate of the column density of HCO$^+$, which is often hampered by the optical thickness of its transitions, especially of its ground state \citep[e.g.][]{vasyunina2012}.

%aim of the paper and IRDCs
Our aim in this work is to minimise these limitations with a new analytical estimate of $\zeta_2$ based on the abundance of N$_2$H$^+$ and N$_2$D$^+$. We will then test this framework at parsec scales in an infrared dark cloud (IRDC). IRDCs were first identified as dark objects that obscure IR stellar light across the Galactic Plane at wavelengths from $\sim10~\mu$m to $100~\mu$m \citep{perault1996}. Systematic surveys confirm IRDCs as extended (sizes $\gtrsim5$~pc) and massive ($\gtrsim5000$~M$_{\odot}$) objects with average densities $\sim2\times10^3$~cm$^{-3}$ \citep{simon2006-cat,simon2006-prop}. High-mass star-formation is often hosted in IRDCs \citep[e.g.][]{peretto2013}, of which the massive ($\gtrsim100$~M$_\mathrm{\odot}$) and compact ($\lesssim0.7$~pc) IR-dark clumps \citep{ragan2006} act as potential precursors \citep{rathborne2006}. Previous works in IRDCs \citep{entekhabi2022} and other Galactic filaments \citep{clarke2024} reported values of $\zeta_2\lesssim4\times10^{-18}$~s$^{-1}$. These ionisation rates, reduced compared to other estimates and theoretical predictions \citep[see][and references therein]{padovani2024}, hinted towards an attenuation of the CR flux due to the density regime of the region or its Galactocentric distance.

\begin{figure*}
    \centering
    \includegraphics[width=0.99\linewidth]{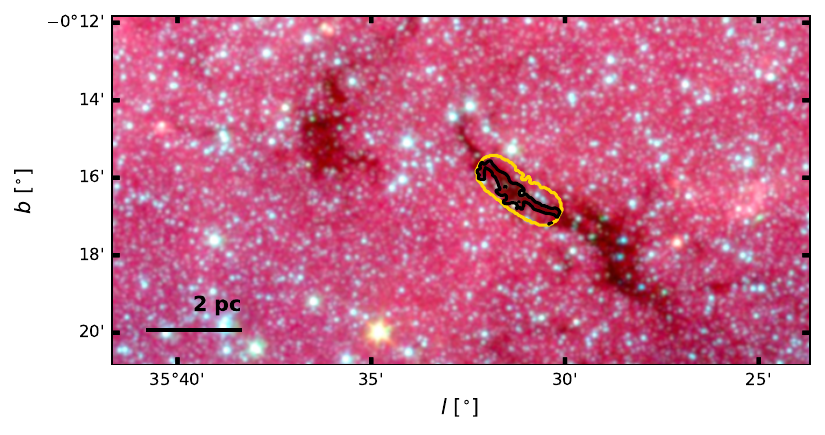}
    \caption{Global MIR view of the IRDC~G035.39-00.33 (Cloud H) within the W48 complex. The background is a three-color \textit{Spitzer} GLIMPSE image (colour-coding RGB: $8~\mu\mathrm{m}$, $5.8~\mu\mathrm{m}$, and $4.5~\mu\mathrm{m}$) \citep{churchwell2009}. The footprint of the NOEMA+IRAM-30m observations targeting the northern filament of Cloud H is overlaid (gold contour), as well as a MIR+NIR extinction level of $A_V=30$~mag, masked to fit the same footprint \citep[black contour;][]{kainulainen2013-maps}. A scale bar of 2~pc is displayed as well.}
    \label{fig:W48}
\end{figure*}

%Cloud H
Our target, G035.39-00.33 \citep[Cloud H hereafter, see][]{butler2009,butler2012} is a massive IRDC ($\sim1.7\times10^4$~M$_{\odot}$) located at a heliocentric distance of 2.9~kpc \citep{simon2006-prop}. Part of the W48 molecular complex \citep[see][and references therein]{luong2011,cosentino2025}, Cloud H appears as a narrow ridge with extinctions $A_V\gtrsim25$~mag \citep{kainulainen2013-maps} engulfed in the feedback from nearby high-mass star-formation sites \citep[see Fig.~\ref{fig:W48} and also][]{liu2026-alma-sio}. This filamentary cloud hosts several clumps and cores with masses $\gtrsim60$~M$_\odot$ \citep{rathborne2006} and up to $\sim30$~M$_\odot$ \citep{butler2012}, respectively, that are potential precursors of stellar clusters. Recent observing campaigns towards Cloud H at low ($\gtrsim20''$) and high ($\lesssim5''$) resolution targeted the dense gas morphology and structure, as probed by N$_2$H$^+$, of its northern filament \citep{henshaw2014,henshaw2016,barnes2018,barnes2021}. These works revealed a complex sub-structure of filaments, fibers, and cores directly connected to the star-formation sites within the fields \citep{carey2009,jimenez-serra2010,luong2011}. Finally, Cloud H exhibits significant degrees of CO depletion \citep[with depletion factors $f_\mathrm{D}\gtrsim4$;][]{hernandez2011,jimenez-serra2014} and high deuterium fractions (probed as $R_\mathrm{D}=\frac{\mathrm{N_2D^+}}{\mathrm{N_2H^+}}\sim0.04$) on scales $\gtrsim1$~pc \citep{barnes2016}. These properties highlight Cloud H as an ideal target for determining $\zeta_2$ at parsec scales with our new analytical framework.

%outline of the paper
The paper is structured as follows: we introduce the method and the main assumptions behind its application (Sect.~\ref{sec:analytical formulae}); we then demonstrate its reliability and limits of application via a direct comparison with the solution from chemical models (Sect.~\ref{sec:comparison with models}); we next introduce the observations, both novel and archival, used in the work (Sect.~\ref{sec:observations}); we continue by presenting the analysis performed and its main results concerning the spatial distribution of the tracers, the determination of the column densities, the degree of depletion, and the deuterium fraction (Sect.~\ref{sec:analysis}); finally, we discuss the electron fraction and the cosmic-ray ionisation rate (CRIR) determined, their correlation with number and column density, respectively, plus their relation to the magnetic field orientation and strength in Cloud H (Sect.~\ref{sec:xe-zeta}). Section~\ref{sec:conclusions} summarises the main findings of this work and our conclusions.

\section{Analytical formulae}\label{sec:analytical formulae}

%Molecular ions: differences between N2H+/N2D+ and HCO+/DCO+
N$_2$H$^+$, HCO$^+$ and N$_2$D$^+$, DCO$^+$ are the direct products of protonation and deuteration of N$_2$ and CO, respectively. The main pathways of formation for these species involve H$_3^+$ and its D-bearing isotopologues (i.e. H$_2$D$^+$, D$_2$H$^+$, D$_3^+$). The properties of these ions and of their precursors, however, deeply influence the gas regime they probe. First, N$_2$ formation in the gas phase is slower than that of CO \citep{aikawa2005} and its freeze-out onto dust grains occurs at higher number densities \citep[a few $\sim10^5$~cm$^{-3}$;][]{caselli2022} compared to the latter species \citep[$n\gtrsim10^4$~cm$^{-3}$;][]{bergin07}. Second, the formation of N$_2$D$^+$ is hindered at temperatures $T_\mathrm{K}\gtrsim20$~K due to decrease of the H$_3^+$ isotopologues in the gas phase \citep{gerlich2002}, while the formation of DCO$^+$ can still occur via CH$_2$D$^+$ \citep{roueff2007,parise2009,oberg2012}. Third, the effective critical densities of N$_2$H$^+$ and N$_2$D$^+$ are higher than those of HCO$^+$ and DCO$^+$ for the same transition \citep{shirley2015}. Finally, the nuclear spin bore by the nitrogen atoms in N$_2$H$^+$ and N$_2$D$^+$ causes the splitting of their spectral lines into multiple hyperfine components \citep{turner1974}. Upon assumption of local thermodynamic equilibrium (LTE), the total intensity of the line, as well as its opacity, is then divided among these components via their statistical weights \citep[although excitation anomalies may exist;][]{caselli95}. Then it is common to find at least one isolated optically thin hyperfine line from which to infer the total intensity \citep[e.g.][]{chen19}, in contrast to the often optically thick HCO$^+$ emission \citep[e.g.][]{vasyunina2012}. These properties allow N$_2$H$^+$ and N$_2$D$^+$ to sample efficiently the cold and dense gas in molecular clouds \citep[i.e. $T\lesssim20$~K, $n\gtrsim10^4$~cm$^{-3}$;][]{bergin07} and make their deuterium fraction $R_\mathrm{D} = \frac{N(\mathrm{N_2D^+})}{N(\mathrm{N_2H^+})}$ a more suitable probe of this gas regime compared to its $\frac{N(\mathrm{DCO^+})}{N(\mathrm{HCO^+})}$ counterpart \citep[e.g.][]{crapsi2005}.

%derivation - part 1: deuterium fraction
The first step in obtaining analytical formulae for the determination of $x(e)$ and $\zeta_2$ is to connect the abundance of H$_3^+$ (i.e. $x(\mathrm{H_3^+})$) to $R_\mathrm{D}$. We consider all relevant formation and destruction pathways related to N$_2$D$^+$ and N$_2$H$^+$. Our chemical network, selected from \citet{sipila2022} (see discussion in Appendix~\ref{sec:fullnetwork}), thus includes at this stage H$_3^+$, all of its isotopologues, all of their spin states (i.e. ortho and para, plus meta for D$_3^+$), and O, CO, electrons ($e^-$), and negatively-charged grains ($g^-$) as main destruction partners. We reduce the size of this network by considering only the reactions whose rate coefficients are $k>10^{-22}$~cm$^3$~s$^{-1}$ at 15~K \citep[typical temperature of IRDCs;][see discussion in Appendix~\ref{subsec:rates}]{pillai2006}. Assuming a steady-state condition for the system (i.e. $\frac{dx}{dt} = 0$, with $x$ molecular abundance relative to H$_2$), we solve the corresponding ordinary differential equations (ODEs) for N$_2$D$^+$ and N$_2$H$^+$. The equivalencies between the rate coefficients allow us to compute the following relation (see Appendix~\ref{subsec:fullderivation}):
\begin{equation}\label{eq:main_Rd}
    n(\mathrm{H_3^+}) = \frac{1}{3}\frac{(1-2R_\mathrm{D})n(\mathrm{H_2D^+}) + (2-R_\mathrm{D})n(\mathrm{D_2H^+}) + 3n(\mathrm{D_3^+})}{R_\mathrm{D}},
\end{equation}
with $n(\mathrm{X}) = n(\mathrm{para-X}) + n(\mathrm{meta-X}) + n(\mathrm{ortho-X})$, where meta- only refers to D$_3^+$. Equation~\ref{eq:main_Rd} has the same form as that of $\frac{N(\mathrm{DCO^+})}{N(\mathrm{HCO^+})}$ \citep[see Eq.~1 in][]{bovino2020}, but now with the deuterium fraction determined via $\frac{N(\mathrm{N_2D^+})}{N(\mathrm{N_2H^+})}$.

%derivation - part 2: electron fraction
One formation pathway of DCO$^+$ is directly related to the three D-bearing isotopologues of H$_3^+$. The corresponding ODE, solved under the steady-state condition, reads as follows:
\begin{align}\label{eq:DCO$^+$}
\begin{split}
    &k_{18}\big[n(\mathrm{H_2D^+}) + 2n(\mathrm{D_2H^+}) + 3n(\mathrm{D_3^+})\big]n(\mathrm{CO}) = \\&n(\mathrm{DCO^+})\big[k_{21}n(e) + k_{22}n(g^-)\big],
\end{split}
\end{align}
with $k_{18}$, $k_{21}$, $k_{22}$ rate coefficients (see Table~\ref{tab:full-network}). By isolating the sum over the three isotopologues of H$_3^+$, we substitute Eq.~\ref{eq:DCO$^+$} into Eq.~\ref{eq:main_Rd} and invert for the electron fraction, which reads:
\begin{equation}\label{eq:xe}
    x(e) = \frac{k_{8}x(\mathrm{CO}) - k_{22}Kx(g^-)}{Kk_{21} - (k_{11}+k_{12})},
\end{equation}
where $k_{11}$, $k_{12}$ and $k_{8}$ are the rate coefficients for the destruction of N$_2$D$^+$ from e$^-$ and CO, respectively, and $K = \frac{k_{20}}{k_{18}}\frac{x(\mathrm{N_2})}{x(\mathrm{CO})}\frac{x(\mathrm{DCO^+})}{x(\mathrm{N_2D^+})}$ with $k_{20}$ being the rate coefficient for the production of N$_2$D$^+$ from N$_2$ and D$_3^+$ (see Table~\ref{tab:full-network}).
Equation~\ref{eq:xe} only assumes steady-state chemistry and requires an informed guess on the abundances of CO and N$_2$, which are much looser constraints compared to previous methods \citep{caselli1998}.

%derivation - part 3: ionisation rate
The derivation of $\zeta_2$ requires additional assumptions instead. We start by considering the interaction of a CR particle with H$_2$. This reaction produces H$_2^+$ with a high branching ratio \citep[i.e. $\gtrsim90\%$;][]{wakelam2024}. H$_2^+$ reacts rapidly with H$_2$ (ortho and para) producing H$_3^+$ (see Appendix~\ref{sec:fullnetwork}). Using the equivalencies among the rate coefficients, the ODE of H$_2^+$ reads:
\begin{equation}\label{eq:h2+}
    n(\mathrm{H_2^+}) = \frac{k_1\zeta_2}{3k_2},
\end{equation}
where $k_1$ is the branching ratio mentioned above and $k_2$ is the rate coefficient for the formation of H$_3^+$ from H$_2^+$. Equation~\ref{eq:h2+} allows the direct determination of $\zeta_2$ once H$_2^+$ is connected to H$_3^+$. We thus derive its ODE (see Eq.~\ref{eq:h3+-full} for the full form):
\begin{align}\label{eq:h3+}
\begin{split}
    &3k_2n(\mathrm{H_2})n(\mathrm{H_2^+}) + \Sigma_\mathrm{o/p}\big[n(\mathrm{H_2D^+})+n(\mathrm{D_2H^+})\big]\big[kn(\mathrm{H_2})\big] = \\&n(\mathrm{H_3^+})\big[\mathrm{O,CO,N_2,HD,D_2},g^-\big]+\\&4n(e)\big[k_9 n(\mathrm{oH_3^+}) + 2k_{10} n(\mathrm{pH_3^+})\big]~,
\end{split}
\end{align}
with the sum running over the ortho and para spin states of H$_2$D$^+$, D$_2$H$^+$, H$_2$, and $\big[\mathrm{O,CO,N_2,HD,D_2},g^-,e^-\big]$ being the contract form of the summation for all reaction rates concerned with the destruction of H$_3^+$. Equation~\ref{eq:h3+} can be simplified with a few assumptions in order to produce an analytic formula: first, the group of reactions o/pH$_2$D$^+$ + o/pH$_2$ is more likely to cause a spin-state change (i.e. o/pH$_2$D$^+\rightarrow$ p/oH$_2$D$^+$) than the production of H$_3^+$ \citep{brunken2014}; second, we assume that the same spin-state change applies for the group of reactions o/pD$_2$H$^+$ + o/pH$_2$; finally, the dissociative recombination of H$_3^+$ due to electrons shows rate coefficients $2k_{10}/k_9\sim9$. Assuming that $n(\mathrm{pH_3^+})\gtrsim n(\mathrm{oH_3^+})$, which is expected in steady state, we neglect the term $k_9 n(\mathrm{oH_3^+})$. Equation~\ref{eq:h3+} then reads:
\begin{align}\label{eq:h3+ simplified}
\begin{split}
    &\zeta_2 k_1n(\mathrm{H_2}) = n(\mathrm{H_3^+})\big[2(k_3+k_4)n(\mathrm{O}) + 2(k_5+k_6)n(\mathrm{CO}) + \\&3k_{20}n(\mathrm{N_2}) +an(\mathrm{HD})+bn(\mathrm{D_2})+ 4k_{13}n(g^-) + 8k_{10}n(e)\big]~,
\end{split}
\end{align}
where $a, b$ are the sum of the rate coefficients for the reactions related to HD and D$_2$ (see Table~\ref{tab:full-network}).
We further assume that $R_\mathrm{D}\lesssim0.15$ to remove it from the numerator of Eq.~\ref{eq:main_Rd} (see Sect.~\ref{subsec:fullderivation}) and combine the latter with Eq.~\ref{eq:DCO$^+$}:
\begin{equation}\label{eq:h3+-final}
    n(\mathrm{H_3^+}) = \frac{1}{3k_{18}}\frac{R_\mathrm{D}^*}{R_\mathrm{D}}\big[k_{21}n(e) + k_{22}n(g^-)\big]~,
\end{equation}
where $R_\mathrm{D}^* = \frac{n(\mathrm{DCO^+})}{n(\mathrm{CO})}$. We finally combine Eq.~\ref{eq:h3+-final} with Eq.~\ref{eq:h3+ simplified} and compute all abundances with respect to $n(\mathrm{H_2})$ obtaining the analytical formula for $\zeta_2$:
\begin{align}\label{eq:zeta}
\begin{split}
    &\zeta_2 = \frac{1}{3k_1k_{18}}\frac{R_\mathrm{D}^*}{R_\mathrm{D}}\big[2(k_3+k_4)n(\mathrm{O}) + 2(k_5+k_6)n(\mathrm{CO}) + \\&3k_{20}n(\mathrm{N_2}) +an(\mathrm{HD})+bn(\mathrm{D_2})+ 4k_{13}n(g^-) + 8k_{10}n(e)\big]\\&\big[k_{21}n(e) + k_{22}n(g^-)\big]n(\mathrm{H_2})~.
\end{split}
\end{align}

%closure: limits of applicability
Equations~\ref{eq:xe} and~\ref{eq:zeta} allow the simultaneous determination of the electron fraction and the CRIR with a rather limited number of assumptions. Nevertheless, these relations still require educated guesses on the abundances of species not probed by the observations and have limitations to their use. The abundances are shown in Table~\ref{tab:model params} and discussed in Appendix~\ref{sec:fullnetwork} where their choice is motivated based on studies from the literature (Sect.~\ref{subsec:abundances-not-probed}) and their values compared to those obtained from the full chemical modelling provided by \texttt{PyRate} (Sect.~\ref{subsec:assumed-model}). The first limitation to the method is that $x(e)$ be positive. Given that $x(\mathrm{CO})\gg x(g^-)$ unless extreme depletion is present, the requirement $x(e)>0$ translates into $Kk_{21} > (k_{11}+k_{12})$, which leads to the following relation (see Eq.~\ref{eq:xe}):
\begin{equation}\label{eq:existence-cond}
    \frac{x(\mathrm{DCO^+})}{x(\mathrm{N_2D^+})}>\frac{k_{18}(k_{11}+k_{12})}{k_{20}k_{21}}\frac{x(\mathrm{CO})}{x(\mathrm{N_2})}\sim2\frac{x(\mathrm{CO})}{x(\mathrm{N_2})}~,
\end{equation}
where the reaction rate ratio is calculated at $15$~K. The second limitation to the method is that $R_\mathrm{D}\lesssim0.15$ to retrieve values of $\zeta_2$ accurate within a factor of 3 (see discussion in Sect.~\ref{subsec:fullderivation}). 

\section{Comparison with chemical models}\label{sec:comparison with models}

%introduction - \texttt{pyRate}
We tested the reliability and range of applicability of our analytical formulation against the chemical modelling provided by \texttt{pyRate} \citep{sipila2015}. \texttt{pyRate} is a 0-D chemical software that constructs and solves a system of ODEs across a defined time range, initial elemental abundances, and initial physical properties. The initial elemental abundances, as well as the chemical network, that we use as input are those from \citet{sipila2022}. The latter comprises prescriptions for the spin-state chemistry, a multi-layered ice chemistry, gas-grain chemistry, plus detailed rate coefficients for the H$_3^+$ + H$_2$ system \citep[see][]{sipila2016,sipila2017,sipila2018}. We will now briefly discuss the reasoning behind the choice of the physical parameters used to run our realisations with \texttt{pyRate}.

\begin{table}
\centering
    \begin{threeparttable}
    \caption{Assumed abundances to compute Eqs.~\ref{eq:xe} and \ref{eq:zeta} (top part) and physical model of the IRDC adopted as framework to run \texttt{pyRate} (bottom part). See Sect.~\ref{sec:comparison with models} and \ref{subsec:abundances-not-probed} for a discussion.}\label{tab:model params}
    \begin{tabular}{c c c}
        \hline\hline
        Parameter & Adopted value & Reference \\ \hline
        \multicolumn{3}{c}{\textit{Assumed abundances}} \\
        $x(\mathrm{HD})$ & $3.2\times10^{-5}$ & a \\
        $x(\mathrm{D_2})$ & $4\times10^{-6}$ & a \\
        $x(\mathrm{N_2})$ & $7.56\times10^{-5}$ & b \\
        $x(\mathrm{O})$ & $2\times10^{-5}$ & c \\
        $x(g^-)$ & $2\times10^{-13}$ & d \\ \hline
        \multicolumn{3}{c}{\textit{Physical model}} \\
        $n(\mathrm{H_2})$ & $10^3-10^6$~cm$^{-3}$ & e, f \\
        $T_\mathrm{K}$ & 15~K & g\\
        $T_\mathrm{dust}$ & 15~K & g \\
        $\zeta_2$ & $1.3\times10^{-17}$~s$^{-1}$ & h \\
        $N(\mathrm{H_2})$ & $10^{22}\Big(\frac{n(\mathrm{H_2})}{2\times10^4~\mathrm{cm}^{-3}}\Big)^{4/3}$~cm$^{-2}$ & i \\ \hline
    \end{tabular}
    \begin{tablenotes}
        \item \textbf{Notes.} References: (a) \citet{flower2004}; (b) \citet{semenov2010}; (c) \citet{vandishoeck2021}; (d) \citet{mckee1989}; (e) \citet{kainulainen2013-maps}; (f) \citet{barnes2021}; (g) \citet{pillai2006}; (h) \citet{webber1998}; (i) \citet{tafalla2021}.
    \end{tablenotes}
    \end{threeparttable}
\end{table}

%\texttt{pyRate} input - physical parameters
Our aim is to estimate $x(e)$ and $\zeta_2$ throughout a cloud, thus spanning a potentially broad range of physical properties. We thus construct a simplified physical model on which we run \texttt{pyRate}. This model is summarised in Table~\ref{tab:model params} and includes the following parameters: a number density range within $n=10^3-10^6$~cm$^{-3}$ sampled in 30 equally-spaced logarithmic bins. The two extrema represent the mean number densities of the IRDCs before they merge with the background material \citep{kainulainen2013-maps} on the one hand and for the dense cores embedded within them on the other \citep{barnes2021}; a constant dust and kinetic temperature fixed at the same value of $T_\mathrm{K} = T_\mathrm{dust} = 15$~K. This value is the average temperature determined in a sample of typical IRDCs using NH$_3$ observations \citep{pillai2006}; a constant CRIR of $\zeta_2=1.3\times10^{-17}$~s$^{-1}$. This is the standard ionisation rate from Galactic CRs determined for H$_2$ in the ISM \citep[see][for a discussion on the ionisation rate for atomic and molecular gas]{glassgold1974,webber1998}; a link between column density and number density, which we identify in the analytical prescription from \citet{tafalla2021} (see Table~\ref{tab:model params}). The latter reproduces well the emission of several molecular tracers, modelled by radiative transfer calculations, across various temperatures and clouds \citep{tafalla2023} and completes a physical model representative of the typical properties expected in G035.39-00.33.

\begin{figure}
    \centering
    \includegraphics[width=0.99\linewidth]{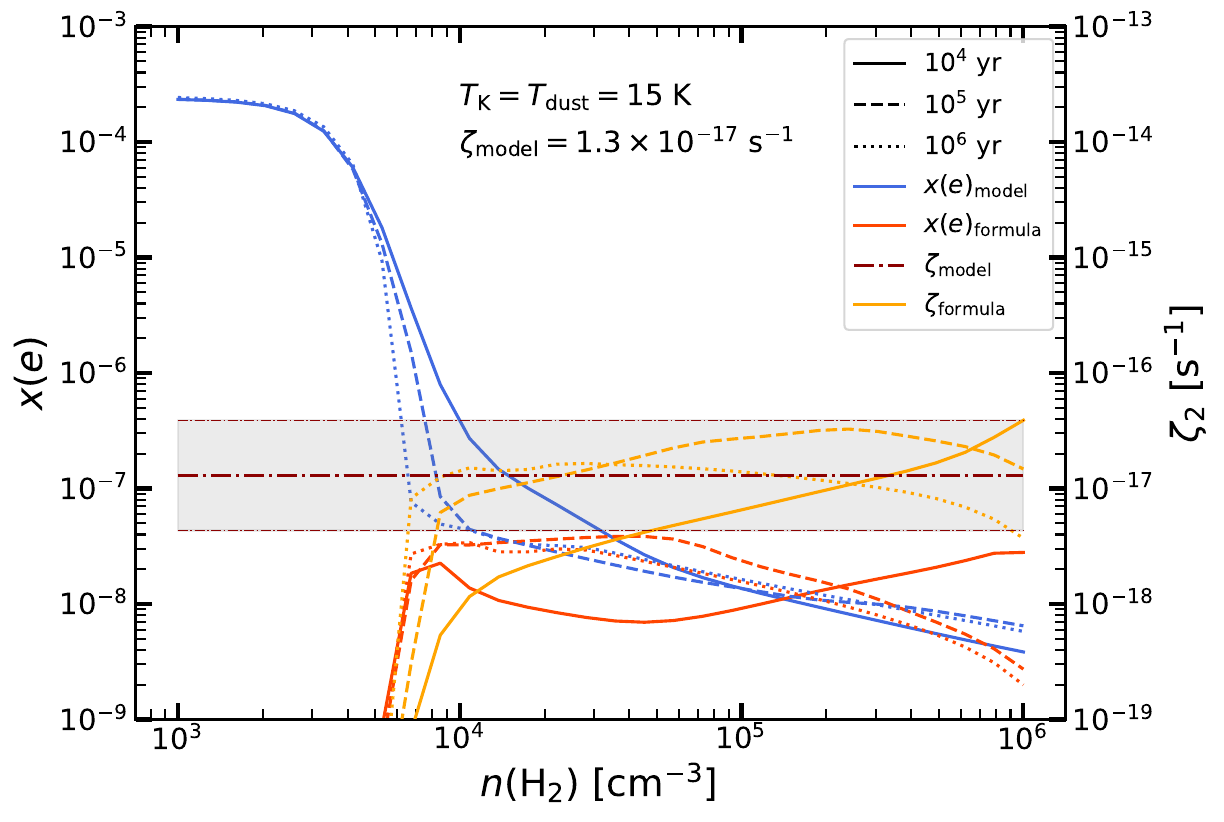}
    \caption{Values of the $x(e)$ and $\zeta_2$ determined at different number densities of H$_2$. The blue and dark red lines represent the electron fraction computed by the model ($x(e)_\mathrm{model}$) and the CRIR input in \texttt{pyRate} ($\zeta_\mathrm{model}$), respectively. The red and orange lines represent instead the electron fraction and the CRIR computed using Eqs.~\ref{eq:xe},~\ref{eq:zeta} and the molecular abundances derived from \texttt{pyRate} ($x(e)_\mathrm{formula}$, $\zeta_\mathrm{formula}$), respectively. The shaded area represents a factor of 3 around $\zeta_\mathrm{model}$. The molecular and atomic abundances were taken at three representative times, namely $10^4$~yr (solid lines), $10^5$~yr (dashed lines), and $10^6$~yr (dotted lines).}
    \label{fig:formulavsmodel}
\end{figure}

%\texttt{pyRate} run - the different regimes
We perform individual runs of \texttt{pyRate} in each of the 30 number density bins included in our model. For each bin, we then extract the number densities of the species of interest ($n(\mathrm{X})$, with X being the chosen molecule) and compute their abundance relative to H$_2$ (i.e. $x(\mathrm{X}) = n(\mathrm{X})/n(\mathrm{H_2})$). Figure~\ref{fig:formulavsmodel} shows the comparison between the electron fractions and the cosmic-ray ionisation rates in each number density bin sampled by the model. The two parameters are taken directly from the model ($x(e)_\mathrm{model}$, $\zeta_\mathrm{model}$, blue and dark red lines, respectively) or calculated from Eqs.~\ref{eq:xe} and \ref{eq:zeta} using the abundances determined with \texttt{pyRate} instead ($x(e)_\mathrm{formula}$, $\zeta_\mathrm{formula}$, red and orange lines, respectively). The electron and molecular abundances are taken from the model at three representative times, these being $10^4$~yr (solid lines), $10^5$~yr (dashed lines) and $10^6$~yr (dotted lines), to ease visualisation. Several different regimes can be observed in Fig.~\ref{fig:formulavsmodel}: for $n\lesssim6\times10^3$~cm$^{-3}$, the values derived from the formulae differ significantly from those of the chemical model. This difference comes from the UV photons driving the ionisation at lower densities with the atomic ions, such as H$^+$ and C$^+$, dominating over the molecular ones. Based solely on the latter, Equations~\ref{eq:xe} and \ref{eq:zeta} cannot efficiently probe this regime in the cloud; for $n\gtrsim10^4$~cm$^{-3}$, and over approximately two orders of magnitude in density, the formulae agree with the chemical model within a factor of 3 in most instances. Departures larger than a factor of 3 are found only when taking abundances computed at $10^4$~yr. This departure, which progressively reduces for later evolutionary times, is a consequence of the steady-state assumption being violated at earlier times. Multiple processes are in fact sensitive to time dependence, with the CO depletion rate \citep[higher at later times for comparable densities; e.g.][]{bergin07} and the abundance of N$_2$ \citep[higher at later times due to its slow formation rate; e.g.][]{aikawa2005} in the lead; for $n\sim10^6$~cm$^{-3}$, $\zeta_2$ still agrees with the input value of the model within a factor of 3, however, it shows a clear downward trend for later times as well. This decrease for later times likely comes from $x(e)$, which shows a similar downward trend that departs from the chemical model calculations. We note that the grain contribution in Eq.~\ref{eq:xe} appears to cause this deviation. The removal of this term from Eq.~\ref{eq:xe} (i.e. $q_2x(g^-)$) produces in fact a much better agreement between the electron fraction $x(e)_\mathrm{formula}$ and $x(e)_\mathrm{model}$ for times later than $10^4$~yr. We interpret this behaviour as an intrinsic balance between the abundance of negatively-charged grains and that of ions not probed through our formulae, such as H$^+$. The latter may remain as abundant in the gas phase as the molecular ones, even for densities $\gtrsim10^6$~cm$^{-3}$ \citep{flower2004}.

%closure: caveats of the method
The chemical models validate our analytical method as most reliable for a range of number densities $n\sim10^4-10^6$~cm$^{-3}$ and for cloud ages of $\gtrsim10^5$~yr. This is consistent with both the steady-state assumption and the assumption of low $R_\mathrm{D}$: at earlier times, the depletion of CO is yet to be effective and N$_2$ is not abundant enough to grant detectable abundances of N$_2$H$^+$ and N$_2$D$^+$; at higher number densities, D$_3^+$ dominates the deuteration processes and the error in the estimate of $\zeta_2$ increases. Number densities $n\sim10^4-10^6$~cm$^{-3}$ correspond to the expected values between cloud ($\gtrsim1$~pc) and core ($\sim0.1$~pc) scales \citep[e.g.][]{bergin07}, thus making our method particularly suited to probe $\zeta_2$ in dense, large-scale filaments.

\begin{figure*}
    \centering
    \includegraphics[width=0.99\linewidth]{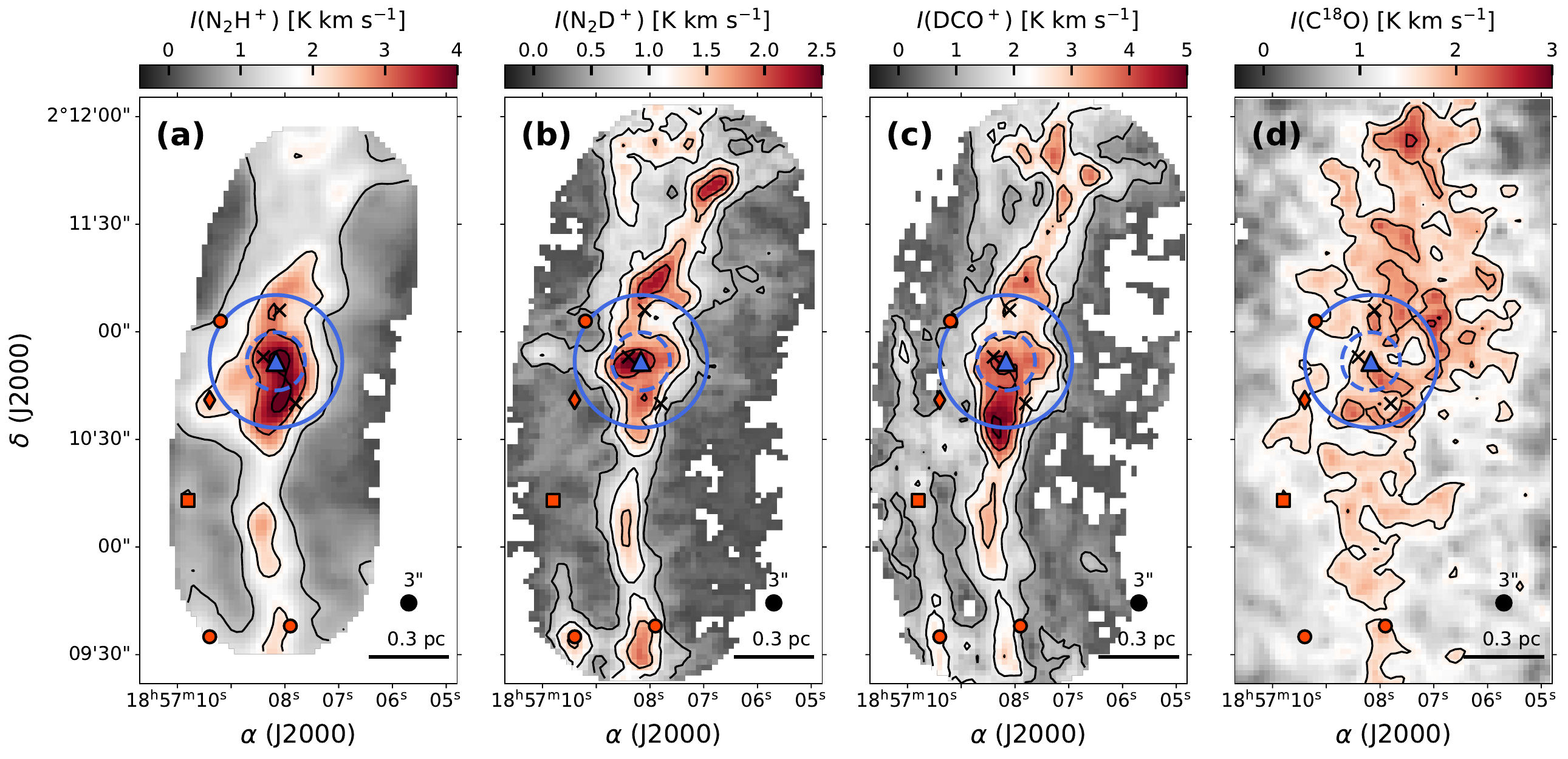}
    \caption{Velocity-integrated intensity maps of the ground transition of N$_2$H$^+$ (panel a), N$_2$D$^+$ (panel b), DCO$^+$ (panel c), and C$^{18}$O (panel d) towards Cloud H. The contours correspond to [1, 2, 3, 4]~K~km~s$^{-1}$ for N$_2$H$^+$, [0.5, 1, 1.5, 2]~K~km~s$^{-1}$ for N$_2$D$^+$, [1, 2, 3, 4, 5]~K~km~s$^{-1}$ for DCO$^+$, and [1.5, 2, 2.5]~K~km~s$^{-1}$ for C$^{18}$O. The blue triangle represents the central position of the H6 core \citep{butler2012} within the MM7 clump \citep{rathborne2006} with their typical sizes reported in these studies displayed as dashed and solid circles, respectively. The red circles, diamond, and square represent the dense pre-stellar cores, the dense core with MIR counterpart \citep{gutermuth2015}, and an unclassified source reported by \citet{luong2011}. The black crosses represent those sources classified as protostars by these same authors.}
    \label{fig:NOEMAmaps}
\end{figure*}

\section{Observations}\label{sec:observations}

%introduction
We aim to test our analytical method in a filament where observations of our key molecular species, namely N$_2$H$^+$, N$_2$D$^+$, DCO$^+$, CO, and information on other physical properties (e.g. kinetic temperature, column density of H$_2$) are available. Our target of choice is the northern filament of the IRDC G035.39-00.33, or Cloud H. The region was targeted with new high-resolution observations (Sect.~\ref{subsec:noemaobs}) which, once further complemented with ancillary observations from various facilities (Sect.~\ref{subsec:ancillary}), will allow us to measure the CRIR via Eq.~\ref{eq:zeta}.

\subsection{NOEMA+30m observations}\label{subsec:noemaobs}

%NOEMA observations
Cloud H was surveyed in the ($J=1-0$) transition of N$_2$D$^+$ (77.109~GHz), N$_2$H$^+$ (93.173~GHz) and DCO$^+$ (72.039~GHz) using interferometric observations at 3mm from the NOrthern Extended Millimetre Array (NOEMA). The observations were carried out in two different sessions on the 13th and 16th of August 2021 in configuration D and on the 16th of March 2022 in configuration C (Project: S21AF001; PI: A.~T.~Barnes) with baselines between 16.4~m and 287.8~m. Both sessions leveraged on 10 antennas and covered the northern filament with a mosaic of four Nyquist-sampled pointings. The nominal resolution achieved by the observations was $2.8''\times2.3''$ in configuration C and $5.1''\times4.7''$ in configuration D at the rest frequency of N$_2$D$^+$. We used the Band 1 receiver and connected to the PolyFiX correlator with the local oscillator (LO) configured at a frequency of $83.5$~GHz. The correlator attained an instantaneous bandwidth of 31.5~GHz at a spectral resolution of 2~MHz. Individual high-resolution chunks with variable bandwidth were configured to cover our tracers of interest at a spectral resolution of 62.5~kHz (or $\sim0.25$~km~s$^{-1}$). A total of 3~hrs in configuration D and 3.7~hrs in configuration C were allocated to achieve a line sensitivity of $\sim5.4$~mJy~beam$^{-1}$ (or $\sim0.15$~K).

%IRAM-30m observations
Complementary single-dish observations at 3mm were requested as short-spacings for the NOEMA visibilities within the same project. These observations were carried out at the 30-metre Instituto de Radioastronomia Mill\'{i}metrica telescope (IRAM-30m) in Granada (Spain) during the summer semester of 2021 (Project: 091-21). Observations with the IRAM-30m were obtained in on-the-fly (OTF) position-switching (PSw) mode and designed to cover the large NOEMA mosaic plus a band-guard for a total area of $\sim1.5\times2.5$~arcmin$^2$. The Eight Mixer Receiver (EMIR) \citep{carter2012} was configured in Band E0 and used as frontend. To preserve the PolyFiX spectral resolution, we connected the Fast Fourier Transform Spectrometer \citep[FTS;][]{klein2012} with a resolution of 50~kHz (FTS50) as the backend. Its instantaneous bandwidth is smaller than that of PolyFiX, thus we required two tunings centred on the frequencies 72.82~GHz and 74.32~GHz, respectively, in the lower outer sub-band. A total of 22~hrs was allocated to achieve a line sensitivity of 15~mK at a resolution of 50~kHz.

%data combination and imaging
We reduced the NOEMA interferometric visibilities using the standard observatory pipeline in the Continuum and Line Interferometer Calibration (CLIC) software part of the Grenoble Image and Line Data Analysis Software (GILDAS) package \citep{pety2005,gildas2013}. The calibrated visibilities of the C and D configurations were then concatenated and imaged in the MAPPING software with custom scripts. The IRAM-30m cubes obtained in antenna temperature ($T_\mathrm{a}^*$) were first converted into main beam temperature ($T_\mathrm{mb}$) through the relation $T_\mathrm{mb}=T_\mathrm{a}^*F_\mathrm{eff}/B_\mathrm{eff}$. Here, we adopted the standard forward ($F_\mathrm{eff}$) and backward ($B_\mathrm{eff}$) efficiencies for the IRAM-30m telescope\footnote{\url{https://publicwiki.iram.es/Iram30mEfficiencies}}: N$_2$D$^+$ (1$-$0): $B_\mathrm{eff}=0.82$, $F_\mathrm{eff}=0.95$; N$_2$H$^+$ (1$-$0): $B_\mathrm{eff}=0.80$, $F_\mathrm{eff}=0.95$; DCO$^+$ (1$-$0): $B_\mathrm{eff}=0.82$, $F_\mathrm{eff}=0.95$. We then applied the continuum subtraction to the interferometric visibilities and merged them with the single-dish pseudo-visibilities using the task \texttt{uv\_short}. The table was imaged using the \texttt{uv\_map} task and the image was cleaned using natural weighting and the Hogbom deconvolver \citep{hogbom1974}. We convolved the NOEMA+30m spectral cubes to a final resolution of $3''$ and applied a grid of three pixels per beam. The absolute flux scale is estimated to be accurate within a $10\%$.

\subsection{Auxiliary observations}\label{subsec:ancillary}

%introduction
In addition to our suite of ions, Eq.~\ref{eq:zeta} requires an estimate of other parameters: the total column (and number) density of H$_2$, $N(\mathrm{H_2})$ ($n(\mathrm{H_2})$), the degree of CO depletion, expressed through the depletion factor $f_\mathrm{D}$ and the kinetic temperature $T_\mathrm{K}$. Thanks to the wealth of data available for Cloud H, we can estimate these parameters with ancillary observations from various facilities.

%N(H2) from FIR+MIR extinction
Figure~\ref{fig:archivaldata} (left) shows the total column density of H$_2$ towards the northern filament of Cloud H. The map was obtained from a cross calibration of the NIR \citep[UKIDSS/GPS photometry in the JHK bands;][]{lucas2008,kainulainen2011} and MIR \citep[$8~\mu\mathrm{m}$ data from \textit{Spitzer}/GLIMPSE;][]{benjamin2003,butler2012} dust extinction measurements at a native resolution of $\sim2''$. We refer the reader to \citet{kainulainen2013-maps} for a full description of the technique and to Sect.~\ref{sec:additional maps} for a description of the filament structure as seen in dust extinction. We briefly note a few caveats to the technique: the MIR data saturates at high column densities, thus becoming unreliable at extinctions $\sim90-115~A_V$ \citep{butler2012}; a column density map based on dust extinction is not applicable in regions containing localised IR sources, which appear as regions of artificially low (or unconstrained) values of $N(\mathrm{H_2})$ \citep{butler2009}; combining the two IR datasets requires knowledge of the dust opacity law, for which the authors report an uncertainty to the method of $36\%$ \citep{kainulainen2013-maps}. The first two limiting factors have a mild impact on our analysis as the highest extinction values in the region are $\sim70~A_V$ and most bright IR sources reside outside the filament spine \citep[see Fig.~\ref{fig:archivaldata}, left panel;][]{kainulainen2013-maps}. The original map, obtained in units of mass surface density ($\Sigma$), was converted to molecules of H$_2$ by assuming solar abundances of the elements \citep{asplund2021} and dividing by $\mu_\mathrm{H_2}m_\mathrm{p}$, with $\mu_\mathrm{H_2}=2.81$ mean particle mass per H$_2$ and $m_\mathrm{p}$ mass of the proton.

%n(H2) from N(H2)
Figure~\ref{fig:archivaldata} (centre) shows the number density of H$_2$ towards the northern filament of Cloud H. The map was obtained by applying a denoising diffusion probabilistic model (DDPM) to the original surface density map of the region. We refer the reader to \citep{xu2023} for a full account on the implementation of the technique to this dataset and to Sect.~\ref{sec:additional maps} for a description of the filament structure in number density. We discuss here a few limitations of the technique: the DDPM retains the resolution of the original map (i.e. $\sim2''$), but it is similarly deficient towards the IR-bright sources; comparisons with full magnetohydrodynamic (MHD) simulations reveal that the DDPM underestimates by $\sim10\%$ the expected density structure of the cloud with a $\sim38\%$ deviation around its mean value. The map, initially reconstructed as total number density, $n = n_\mathrm{H}+2n(\mathrm{H_2})$, was converted in density of H$_2$ by assuming a negligible fraction of atomic hydrogen (i.e. $n\sim2n(\mathrm{H_2})$).

%C18O data
Figure~\ref{fig:NOEMAmaps} (panel d) shows the velocity-integrated intensity of C$^{18}$O (1$-$0) towards the northern filament of Cloud H. The observations were performed on the 29th of March 2022 (PI: J. Tan) using the 16-element heterodyne radio camera Argus \citep{sieth2014} mounted on the Green Bank Telescope (GBT). The map was obtained using the fast-mapping technique by scanning the pixel receivers along Galactic longitude in rows sampled every $5.58''$. After the baseline removal and re-gridding, the final C$^{18}$O (1$-$0) spectral cube has a pixel scale of $2.57''$ and a spectral resolution of 0.182~km~s$^{-1}$. The rms noise level, determined across channels without emission, is 0.55 K with an absolute flux calibration error of $15\%$. We refer the reader to \citet{chiyan2024} for a full account on the data acquisition and reduction \citep[performed in G28.37+0.07, or Cloud C,][but with the same observational setup and calibration as for Cloud H]{butler2009}, while the C$^{18}$O emission is further compared to that of the ions in the next section.

%determination of Tk
Figure~\ref{fig:archivaldata} (right) shows the kinetic temperature map towards the northern filament of Cloud H. The map was obtained with observations from the Very Large Array (VLA) of the first two inversion transitions of ammonia, NH$_3$ (1,1) and (2,2) \citep{sokolov2018}. These data were further combined with ancillary GBT observations of the same transitions \citep{sokolov2017} to recover the total flux and achieve a final resolution of $5.44''$. The kinematics of Cloud H is complex, and its structure comprises multiple velocity components along the line of sight \citep[e.g.][]{henshaw2013}. Our kinetic temperature map is derived from the modelling of the emission from NH$_3$ (1,1) and (2,2) towards the velocity-component F2 \citep{sokolov2019}, which encompasses the full coverage of our NOEMA+IRAM-30m map. We refer to \citet{sokolov2019} for details on the analysis of the ammonia transitions and to Sect.~\ref{sec:additional maps} for a description of the temperature variations across the filament.

%closure
These ancillary observations provide the final ingredients for measuring the electron fraction and CRIR throughout the cloud. To achieve a more direct comparison and given the similar resolution among all observations, these ancillary maps were re-gridded to the same resolution and pixel size of our NOEMA maps. In the next section, we will make use of these observations to study the link between our molecular tracers, the main parameters derived from their analysis, and their contribution in Eqs.~\ref{eq:xe} and \ref{eq:zeta}, along with those of density and temperature. 

\begin{figure*}
    \centering
    \includegraphics[width=0.99\linewidth]{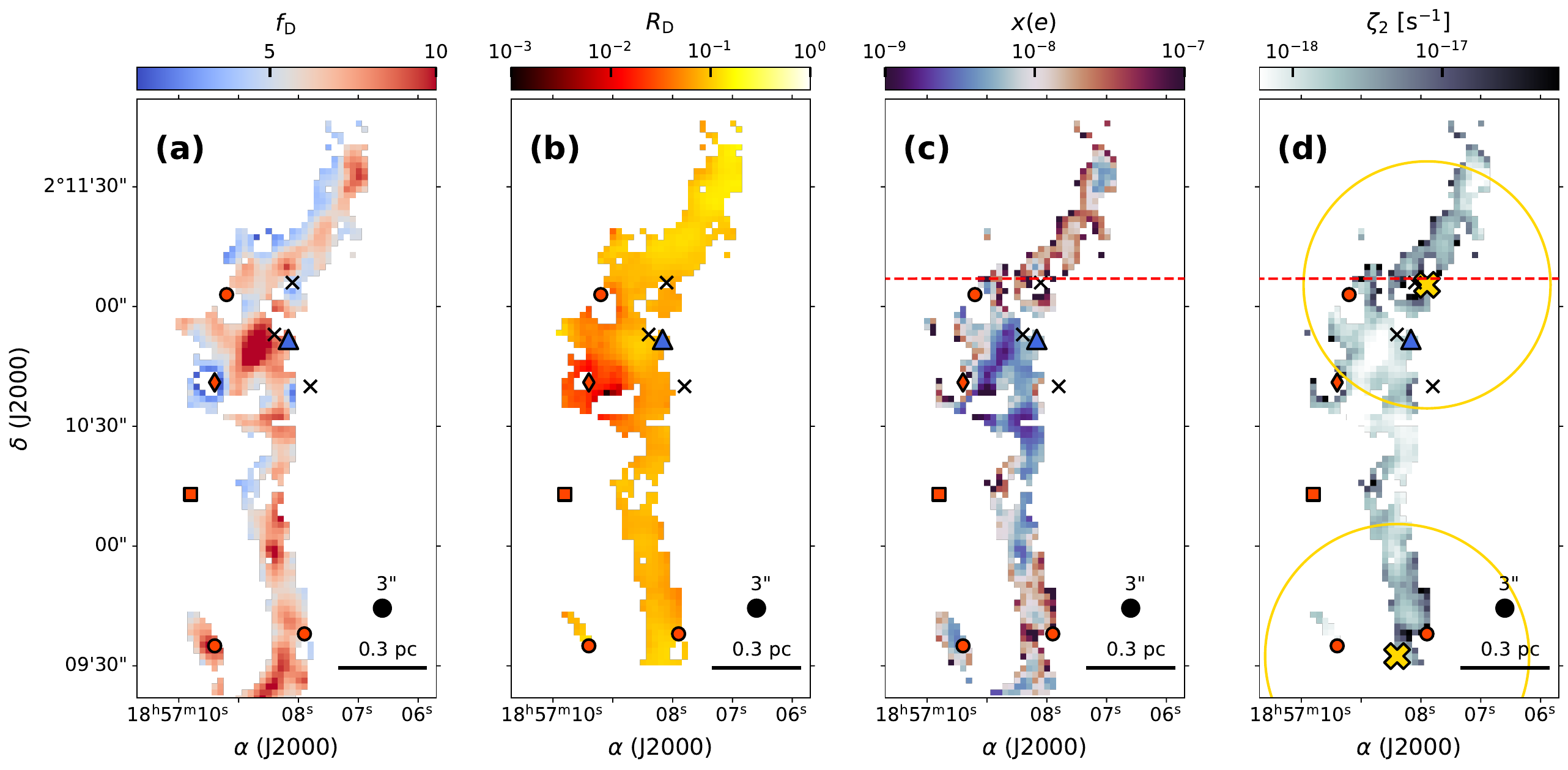}
    \caption{Main results obtained from the analysis of Cloud H (see Sect.~\ref{sec:analysis}). \textit{Panel a}: depletion factor, $f_\mathrm{D},$ derived from the ratio of the expected C$^{18}$O abundance in Cloud H, $x_0$, and the measured abundance, $N(\mathrm{C^{18}O})/N(\mathrm{H_2})$ (see Sect.~\ref{subsec:fd}). \textit{Panel b}: deuterium fraction derived as $R_\mathrm{D}=N(\mathrm{N_2D^+})/N(\mathrm{N_2H^+})$ (see Sect.~\ref{subsec:Rd}). \textit{Panel c}: electron fraction, $x(e)$, determined via Eq.~\ref{eq:xe} in our analytical framework. \textit{Panel d}: CR ionisation rate, $\zeta_2$, determined via Eq.~\ref{eq:zeta} in our analytical framework. The position of the clumps identified by \citet{liu2018} is marked with crosses along with their effective size (gold line). The red dashed lines in the last two panels represents the approximate declination at which the dominant magnetic field component changes \citep[from poloidal towards the North, to toroidal towards the South;][]{liu2018}. The symbols in all panels are the same as in Fig.~\ref{fig:NOEMAmaps}.}
    \label{fig:results}
\end{figure*}

\section{Physical properties of Cloud H}\label{sec:analysis}

%intro
Employing the observations just described, we compute the CRIR throughout Cloud H using Eq.~\ref{eq:zeta}. The parameters needed to perform this calculation are described as follows: first, we discuss the spatial morphology of our suite of molecular tracers (Sect.~\ref{subsec:morphology-emission}); second, we use the integrated intensity maps of our tracers to determine their column densities under a few assumptions (Sect.~\ref{subsec:columns}); third, we determine the CO depletion factor, $f_\mathrm{D}$, from the C$^{18}$O column density (Sect.~\ref{subsec:fd}), the deuterium fraction, $R_\mathrm{D}$, from the N$_2$D$^+$ and N$_2$H$^+$ column densities (Sect.~\ref{subsec:Rd}), and discuss their values in the context of the cloud; finally, we compute the electron fraction, $x(e)$ (Sect.~\ref{subsec:xe-result}), and the CRIR, $\zeta_2$ (Sect.~\ref{subsec:CRIRcloudH}), using Eqs.~\ref{eq:xe}, and \ref{eq:zeta}, respectively.

\subsection{Spatial morphology of the molecular tracers}\label{subsec:morphology-emission}

%3rd paragraph - description of the data
Figure~\ref{fig:NOEMAmaps} shows the combined NOEMA+30m velocity-integrated maps for our suite of ions, paired with the GBT velocity-integrated map for C$^{18}$O. The spectral cubes of DCO$^+$ and N$_2$D$^+$ were integrated over the velocity range $\Delta V_\mathrm{lsr} = [40,~50]$~km~s$^{-1}$ around the typical velocity of Cloud H \citep[i.e. 44.7~km~s$^{-1}$;][]{rathborne2006}. The spectral cube of N$_2$H$^+$ was integrated in the velocity range $\Delta V_\mathrm{lsr} = [30,~40]$~km~s$^{-1}$ around its isolated hyperfine component \citep[i.e. $JF_1F\rightarrow J'F'_1F' = 101\rightarrow 012$;][]{caselli95}. The choice is driven by the optical depth of the main hyperfine group across Cloud H, which could bias the determination of the deuterium fraction (see Sect.~\ref{subsec:columns} for the full discussion). N$_2$H$^+$ (panel a), N$_2$D$^+$ (panel b) and DCO$^+$ (panel c) all show extended emission arranged in a filamentary shape which closely follows the northern filament seen in total column density (see Sect.~\ref{subsec:ancillary}). Despite this close similarity in the morphology at parsec scales, the three tracers show differences when looking at the sub-parsec structures. N$_2$H$^+$ and N$_2$D$^+$ show their emission peak towards the most massive core within the filament, H6 \citep[$\sim60$~M$_\odot$, blue triangle and dashed circle;][]{butler2012} embedded within the MM7 clump \citep[solid circle;][]{rathborne2006}. The intensity of N$_2$H$^+$ shows a correlation with dense cores and protostars \citep[red circles and black crosses, respectively;][]{luong2011}, while for N$_2$D$^+$ such correlation is lost. The latter shows a more fragmented appearance compared to N$_2$H$^+$ with several of its intensity peaks (i.e. $I\gtrsim1.5-2$~K~km~s$^{-1}$) anti-correlated to those of its main isotopologue and to positions of cores and protostars. The only exception is the protostar towards H6, which falls within the brightest N$_2$D$^+$ peak. This object has no mid-infrared (MIR) counterpart to its emission \citep[i.e. no $8\mu$m and $24\mu$m emission;][]{gutermuth2015}, suggesting a very young protostar with a substantial envelope \citep{luong2011}. DCO$^+$ shows an even different emission pattern. Its peak intensity is within the MM7 clump as well, but shifted towards the South compared to the H6 core position. Its emission above $\gtrsim3$~K~km~s$^{-1}$ is fragmented in scattered peaks which only marginally superpose with those of N$_2$H$^+$ and N$_2$D$^+$. Finally, DCO$^+$ does not show any obvious correlation with the dense cores and protostars in the field.

%morphology of C18O
The morphology of the C$^{18}$O emission is radically different from that of ionic species (panel d) and $N(\mathrm{H_2})$ in Cloud H (black contours). Its intensity peaks (i.e. $I(\mathrm{C^{18}O})\sim3$~K~km~s$^{-1}$) are scattered and usually reside outside the higher column density contours. Moreover, these C$^{18}$O peaks have intensities only a factor of $\sim2-3$ higher than the more diffuse component (which sits at $I(\mathrm{C^{18}O})\sim1-1.5$~K~km~s$^{-1}$). The morphology of its emission therefore appears flat and featureless compared to that of molecular ions and to the column density of H$_2$. This systematic differentiation was first observed on core scales \citep[$\sim0.1$~pc;][]{willacy1998,caselli1999,tafalla2002} and recently discussed on parsec scales in Orion \citep{socci2024-networks}. The latter authors ascribe this decoupling of the CO emission from the column density peaks as a symptom of depletion already occurring on cloud scales. Depletion factors of $f_\mathrm{D}\sim3$ measured on scales $\gtrsim1$~pc in Cloud H support this interpretation \citep{hernandez2011,jimenez-serra2014}.

%closure
The various morphologies of the emission from our tracers hint towards a physical differentiation in Cloud H. N$_2$D$^+$ and DCO$^+$ show emission peaks often anti-correlated to those of N$_2$H$^+$ and within themselves as well. These spatial changes in emission could imply a shift in the dominant deuteration process that occurs within the cloud (see Sect.~\ref{sec:analytical formulae}) due to external factors, such as temperature, density, and the CRIR.

\subsection{Molecular column densities}\label{subsec:columns}

%how to compute the column density
Our survey comprises four molecular tracers, N$_2$H$^+$, N$_2$D$^+$, DCO$^+$, C$^{18}$O, for which we observed their ground transition. We will therefore use their integrated intensities, $I$ (see Fig.~\ref{fig:NOEMAmaps} and \ref{fig:archivaldata} central panel), to derive their column densities as follows \citep{mangum2015}:
\begin{equation}\label{eq:columndenscal}
    N = \frac{8\pi\nu^3}{c^3 g_uA_{ul}}\frac{ Q_\mathrm{rot}}{J_\mathrm{ex} - J_\mathrm{bg}}\frac{I}{1 - e^{-h \nu/k_\mathrm{B}T_\mathrm{ex}}}e^{-E_l/k_\mathrm{B}T_\mathrm{ex}},
\end{equation}
where $c$ is the speed of light, $k_\mathrm{B}$ is the Boltzmann constant, $\nu$ is the rest frequency of the $u\rightarrow l$ transition, $A_{ul}$ is its Einstein coefficient for spontaneous emission, $g_u$ the upper state degeneracy of the transition, $Q_\mathrm{rot}$ is the partition function, $J$ is the black body emission function at the excitation temperature, $T_\mathrm{ex}$ ($J_\mathrm{ex}$), and at the background temperature, $T_\mathrm{bg}=2.725$~K ($J_\mathrm{bg}$). 

%optical depth: N2H+
Equation~\ref{eq:columndenscal} assumes that the emission is optically thin (i.e. with an optical depth $\tau\ll1$), which may not be true for all tracers. DCO$^+$ and N$_2$D$^+$ show optically thin emission even in regions with column densities up to $\sim10^{23}$~cm$^{-2}$ \citep[e.g.][]{sabatini2020,redaelli2022}, therefore we will assume that their emission is optically thin in Cloud H as well. The same cannot be said as easily for N$_2$H$^+$ and C$^{18}$O. N$_2$H$^+$ displays multiple velocity components throughout Cloud H and its emission has been found to be moderately thick \citep{henshaw2014,barnes2018}. Given that N$_2$D$^+$ shows only one velocity component in the region, and to mitigate the effect of optical thickness, we integrated N$_2$H$^+$ around its isolated and optically-thin hyperfine component ($JF_1F\rightarrow J'F'_1F' = 101\rightarrow 012$). We further determined the column density of N$_2$H$^+$ by scaling Eq.~\ref{eq:columndenscal} with the statistical weight of this isolated component \citep[i.e. 1/9; see][for a discussion of the approach and its limitations]{henshaw2014}.

%optical depth: C18O
C$^{18}$O was reported to have optical depths up to $\tau\sim1$ towards MM7/H6 by \citet{hernandez2011}. These optical depths were determined by the authors via the equation:
\begin{equation}\label{eq:hernandez}
    T_\mathrm{B} = (J_\mathrm{ex} - J_\mathrm{bg})(1 - e^{-\tau}),
\end{equation}
where $T_\mathrm{B}$ is the brightness temperature of the line. The excitation temperature introduced in Eq.~\ref{eq:hernandez} was obtained using standard conversion factors for the column density of C$^{18}$O \citep{wilson1994,lacy1994} and was constrained to a typical value of $\sim7.5$~K \citep{hernandez2011}. However, the high-resolution temperature map from \citet{sokolov2019} shows a significant variation in temperature across Cloud H (see Fig.~\ref{fig:archivaldata}, right panel). Given the low critical density of the C$^{18}$O (1$-$0) line (i.e. $\sim10^3$~cm$^{-3}$), we decided to derive its excitation temperature pixel-by-pixel from the kinetic temperature map (Fig.~\ref{fig:archivaldata}, right panel) and explore its potential influence on the optical depth. $T_\mathrm{ex}$ depends on $T_\mathrm{K}$, assuming a two level system, as follows \citep{goldsmith1999}:
\begin{equation}\label{eq:gold}
    T_\mathrm{ex} = T_\mathrm{K}\bigg[1 + \frac{k_\mathrm{B}T_\mathrm{K}}{h\nu}\mathrm{ln}(1 + A_{10}/C_{10})\bigg]^{-1},
\end{equation}
where $h$ is the Planck constant, $A_{10}$, $C_{10}=c_{10}n(\mathrm{H_2})$ are the Einstein coefficient for spontaneous emission and the collisional rate for the ground transition, respectively. The latter depends on the collisional coefficient ($c_{10}$, taken together with $A_{10}$, from the LAMDA database\footnote{\url{https://home.strw.leidenuniv.nl/~moldata/}}) and on the number density sampled by our high-resolution map (Fig.~\ref{fig:archivaldata}, central panel).

%the excitation temperature
The excitation temperature determined through Eq.~\ref{eq:gold} is close to thermalisation with values within $\sim7-20$~K and a median of $\sim12$~K. Taking the latter value as representative for Cloud H, an optical depth of $\sim1$ would decrease to $\sim0.3$ for the same brightness temperature. An optical depth of $\sim0.3$ is in agreement with the findings of \citet{jimenez-serra2014} who used observations at a higher resolution ($\sim14''$ vs $\sim22''$) and who assumed a higher excitation temperature (15~K vs 7.5~K) compared to \citet{hernandez2011}. The column density computed with $\tau\sim0.3$ would change from its optically thin limit by $\sim15\%$, well within the uncertainty of our method. We therefore assumed an optically thin emission for C$^{18}$O (1$-$0) and adopted Eq.~\ref{eq:gold} to derive the excitation temperature of our suite of tracers, including the molecular ions. The latter excitation temperatures are within $\sim3.5-6$~K, in close agreement with the previous values derived in the region \citep{henshaw2013,barnes2016}. We are thus confident in using Eq.~\ref{eq:columndenscal} to compute the column densities of our tracers in the optically thin limit.

\subsection{Large-scale CO depletion at high resolution}\label{subsec:fd}

%how to compute fd
The freeze-out of CO onto dust grain surfaces is the main cause of its disappearance from the gas phase. The degree to which this effect influences its observed abundance is estimated with the depletion factor $f_\mathrm{D}$. The latter can be evaluated as the ratio between the expected abundance of CO ($x_0$) and its observed abundance ($x_\mathrm{obs}$), that is, $f_\mathrm{D} = x_0/x_\mathrm{obs}$. We determined $x_0$ via the Galactic scaling relation for C$^{18}$O \citep{giannetti2017}:
\begin{equation}\label{eq:galgradient}
    x_0 = \frac{9.5\times10^{-5}\times 10^{\alpha(D_\mathrm{GC} - D_\mathrm{GC,\odot})}}{^{16}\mathrm{O}/^{18}\mathrm{O}},
\end{equation}
where $D_\mathrm{GC,\odot}=8.15\pm0.15$~kpc is the Galactocentric distance of the Sun \citep{reid2019}, $\alpha=-0.077\pm0.019$~dex~kpc$^{-1}$ is the scaling of the C/H Galactic gradient \citep{mendez2022}, $^{16}\mathrm{O}/^{18}\mathrm{O} = (58.8\pm11.8)D_\mathrm{GC}+(37.1\pm82.6)$ is the O isotopic ratio across the Milky Way \citep{wilson1994}. This scaling and Eq.~\ref{eq:galgradient} were calculated taking $D_\mathrm{GC}=6.27$~kpc, Galactocentric distance of Cloud H \citep[see][for a discussion]{rathborne2006}. We determined $x_\mathrm{obs}$ taking the ratio of $N(\mathrm{C^{18}O})$, estimated as discussed in Sect.~\ref{subsec:columns}, over $N(\mathrm{H_2})$ (Fig.~\ref{fig:archivaldata}, left panel).

%the depletion map
The map of CO depletion factors shows values $f_\mathrm{D}\gtrsim3$ in the majority of Cloud H (see Fig.~\ref{fig:results}, panel a). This widespread depletion, already reported by \citet{hernandez2011}, is confirmed in our maps with a factor of $\sim7$ better resolution. Both studies recover the highest depletion towards MM7, a milder one in localised spots along the filament spine, and the lowest one towards point sources and protostars, where both the local heating from these objects \citep{draine2011} and the artificially low column density (see Sect.~\ref{subsec:ancillary}) can produce a lower $f_\mathrm{D}$. \citet{hernandez2011} reports CO depletion factors, excluding the cloud envelope (see discussion in their paper), $f_\mathrm{D}'\sim1-5$. Even considering an additional $\sim10-50\%$ factor for the more diffuse CO component, we report degrees of depletion a factor of $\sim2$ higher with $f_\mathrm{D}\sim1-13$ throughout the map and a corresponding median and interquartile range of $\langle f_\mathrm{D}\rangle=6.5^{+1.2}_{-1.0}$. These values are in closer agreement with the depletion factors of $\sim5-12$ reported by \citet{jimenez-serra2014}, with the highest values again found towards the MM7/H6 region. Given the similar sites of enhanced depletion, or higher $f_\mathrm{D}$, reported between the three studies, we can ascribe the difference in absolute values mainly to the change in resolution ($22''$, or $\sim0.3$~pc; $14''$, or $\sim0.2$~pc; $3''$, or $\sim0.04$~pc) which dilutes $f_\mathrm{D}$ within the larger beam, as proved by simulations \citep{bovino2019}.

\begin{figure}
    \centering
    \includegraphics[width=0.99\linewidth]{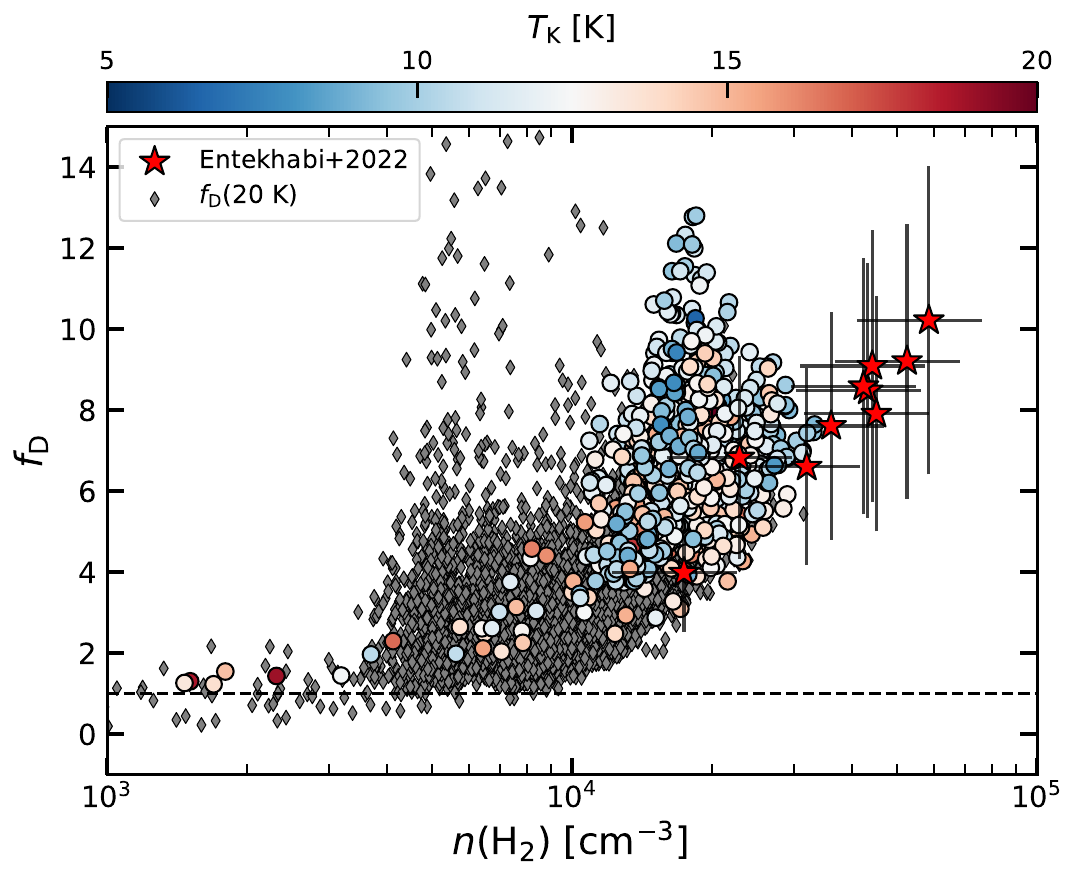}
    \caption{$f_\mathrm{D}$, measured using the kinetic temperature map (Fig.~\ref{fig:archivaldata}), right; circles) and assuming a constant temperature of 20~K (diamonds), against $n(\mathrm{H_2})$ in Cloud H. The two parameters are reported, with errors, for the ten positions sampled in Cloud C as well \citep[stars;][]{entekhabi2022}. The dashed line represents the limit of depletion (i.e. $f_\mathrm{D}=1$).}
    \label{fig:fd-vs-nh2}
\end{figure}

%the relation fd vs n(h2)
A recent study in IRDCs reported CO depletion factors only mildly depending on the number density in the region, while having a negative correlation to its dust temperature instead \citep{cosentino2026}. Figure~\ref{fig:fd-vs-nh2} shows the comparison between depletion factor and number density in Cloud H. When considering those fields with an associated value of $T_\mathrm{K}$, we report a tentative increase in the degree of depletion with the number density. However, the scatter above $n(\mathrm{H_2})\sim10^4$~cm$^{-3}$ is significant and only a few pixels below this density show a positive correlation between the two parameters. We sampled the depletion factors with no associated temperature by computing the C$^{18}$O column density with $T_\mathrm{ex}$ derived from Eq.~\ref{eq:gold} assuming a constant $T_\mathrm{K}=20$~K (value at the map edges, see Fig.\ref{subsec:ancillary}, right). These estimates populate the density regime $n(\mathrm{H_2})\lesssim10^4$~cm$^{-3}$ and the combined distribution of depletion factors obtained with the two estimates shows a positive correlation with the number density across Cloud H. This distribution follows a similar trend to that reported by \citet{entekhabi2022} in Cloud C and is intuitively understood considering the relation between freeze-out timescale and density \citep[i.e. $\tau_\mathrm{fo}\propto n(\mathrm{H_2})^{-1}$;][]{bergin07}. On the other hand, the negative correlation between depletion factor and temperature reported by \citet{cosentino2026} is poorly constrained in Cloud H.
%These results suggest that, while other IRDCs may achieve quasi-equilibrium between CO freeze-out and desorption \citep{cosentino2026}, the latter is less efficient in Cloud C and Cloud H. This conclusion is supported by the low temperatures ($\lesssim15$~K) and cosmic-ray ionisation rates ($\lesssim4\times10^{-18}$~s$^{-1}$) identified in the two regions \citep[][Sect.~\ref{sec:xe-zeta}]{entekhabi2022} which hinder thermal and non-thermal evaporation processes, respectively.

\subsection{Influences on the deuterium fraction in Cloud H}\label{subsec:Rd}

%the range of deuteration values
The deuterium fraction, calculated as $R_\mathrm{D} = N(\mathrm{N_2D^+})/N(\mathrm{N_2H^+})$, shows values ranging two orders of magnitude throughout Cloud H (see Fig.~\ref{fig:results}, panel b). Values within $\sim0.0015-0.15$ with a corresponding median and IQR of $\langle R_\mathrm{D}\rangle=0.08^{+0.02}_{-0.02}$ are consistent with those reported in independent studies of other IRDCs \citep{miettinen2011,gerner2015}. These values are also consistent with those reported in low-mass cores with a similar range in depletion factors and number densities \citep{crapsi2005}. Simulations of turbulent magnetised filaments, such as Cloud H \citep{henshaw2014,liu2018}, report deuterium fractions of $\sim0.01-0.1$ on timescales of $\lesssim0.4$~Myr \citep{koertgen2018}, relatively consistent with our findings. Nevertheless, our values of $R_\mathrm{D}$ remain a factor of a few below those reported in high-mass starless cores \citep[i.e. $R_\mathrm{D}\gtrsim0.3$][]{fontani2011,kong2015}, which is potentially related to the different conditions of those cores compared to Cloud H.

%the variation of Rd across Cloud H
The wide range of deuterium fractions derived in Cloud H is unevenly distributed on the map. Although most of the values along the filament spine cluster around $R_\mathrm{D}\sim0.1$, three main features emerge: a peak towards the North, a peak towards the South, and a dip at the centre. The two peaks show values of $\gtrsim0.1$ extended over sizes of $\sim0.1-0.3$~pc and corresponding to depletion factors around $\gtrsim7$. These high deuterium fractions are co-spatial with the peaks of N$_2$D$^+$ emission and are consistent with the values previously reported at lower resolution in the region \citep{barnes2016}. The dip is centred on the point source with the MIR counterpart located SE of the MM7 clump \citep[red diamond;][]{gutermuth2015}. Here, despite the presence of depletion factors $f_\mathrm{D}\sim10$, the deuterium fraction rapidly decreases from $\sim0.1$ to $\lesssim0.05$ and below. This rapid drop, milder in \citet{barnes2016} due to the lower resolution of the observations (i.e. $27''$), is likely related to the more evolved stage of the point source (see Fig.~\ref{fig:archivaldata}, right).

%the temperature and evolution effects
Deuterium fractions are sensitive to temperature variations as deuteration processes have reaction rates which vary significantly with temperature \citep{watson1974}. The formation of N$_2$D$^+$ from N$_2$ is quite affected by this effect, especially for $T_\mathrm{K}\gtrsim20$~K \citep[e.g.][]{fontani2011,pazukhin2023}. Thus, the aforementioned drop in deuterium fraction towards the centre of the map may be consistent with the increase in temperature (above $\gtrsim15$~K) seen towards the edges of the MM7 clump (see Fig.~\ref{fig:archivaldata}, right panel). The presence of multiple point sources, among which the more evolved one (red diamond), provides a source of gas heating and reinforces the picture of $R_\mathrm{D}$ decreasing from young and starless regions to more star-forming ones reported by previous authors \citep{emprechtinger2009,fontani2011,gerner2015,sabatini2024}.

\section{Electron fraction and cosmic-ray ionisation rate}\label{sec:xe-zeta}

%overview
In the last section, we obtained all the parameters required to determine $x(e)$ and $\zeta_2$ in Cloud H. Now, we will discuss their estimated values as follows: first, we will present the typical values retrieved in Cloud H and their corresponding degree of accuracy (Sect.~\ref{subsec:error-bud}); second, we will outline the typical values and distribution of $x(e)$ (Sect.~\ref{subsec:xe-result}) and $\zeta_2$ (Sect.~\ref{subsec:CRIRcloudH}) across the filament; then, we will address the dependence of $x(e)$ on $n(\mathrm{H_2})$ (Sect.~\ref{subsec:dynamic}); finally, we will discuss the attenuation of $\zeta_2$ with $N(\mathrm{H_2})$ in comparison with the theoretical predictions for the propagation of CRs in the cloud (Sect.~\ref{subsec:padovaniplot}).

\subsection{Analytical estimates and error budget}\label{subsec:error-bud}

%intro: error budget and assumptions
Equations~\ref{eq:xe} and \ref{eq:zeta} allow us to estimate $x(e)$ and $\zeta_2$ with a nominal accuracy of a factor of 3 associated to the method. Many of the parameters embedded in this calculation, however, bear uncertainties: the flux calibration in our observations (Sect.~\ref{subsec:noemaobs}); the systematics in the methods to derive $N(\mathrm{H_2})$ and $n(\mathrm{H_2})$ (Sect.~\ref{subsec:ancillary}); the errors in Galactocentric distance of the Sun and in the Galactic gradients (Sect.~\ref{subsec:fd}). The propagation of these uncertainties can easily grow above the measured values for $x(e)$ and $\zeta_2$. Given that both quantities are definite positive (i.e. with 0 as a lower bound), this propagation is not reasonable nor accurate as it assumed an underlying Gaussian distribution for the two parameters. We therefore assume $x(e)$ and $\zeta_2$ to follow a log-normal distribution, propagate their errors accordingly and report their median values and prediction intervals.

%values for x(e) and zeta
We determined the median and first prediction interval for the full sample of $x(e)$ and $\zeta_2$ measured in Cloud H as $\langle a\rangle=e^{\mu\pm\sigma}$, where $\mu$ and $\sigma$ are the mean and standard deviation, respectively, of the logarithms for the two parameters. Their values correspond to $\langle x(e)\rangle=1.0^{+2.2}_{-0.7}\times10^{-8}$ and $\langle\zeta_2\rangle=2.3^{+5.5}_{-1.6}\times10^{-18}$. Both prediction intervals range around a factor of 3 around their median, thus demonstrating that our data allows us to retrieve average properties for $x(e)$ and $\zeta_2$ with an accuracy consistent with that of the method. Beyond the typical values in Cloud H, we are also interested in studying the local variations of electron fraction and ionisation rate across the filament (Sects.~\ref{subsec:dynamic} and \ref{subsec:padovaniplot}). We therefore propagated the uncertainties in Eqs.~\ref{eq:xe} and \ref{eq:zeta} as sum of Gaussian errors and took them as a pixel-by-pixel deviation ($\sigma$) around each estimate. These errors vary across Cloud H, but their typical magnitude corresponds to a factor of $\sim2$ for $x(e)$ and a factor of $\sim9$ for $\zeta_2$.

%closure: validation of the method
Our estimates return values mostly consistent with the uncertainty associated with the analytical method. Local values can, however, exceed this error, especially in the case of $\zeta_2$. This increase in uncertainty is not surprising given the higher number of independent parameters that contribute to the estimate of $\zeta_2$, and to its error budget, compared to that of $x(e)$ (see Eqs.~\ref{eq:xe} and \ref{eq:zeta}). These high errors, however, do not detract from the quality of the method (if anything, they prompt less uncertain estimates of the parameters fed into $\zeta_2$) and will have only a marginal impact on the conclusions of our work (Sects.~\ref{subsec:padovaniplot} and \ref{sec:conclusions}).

\subsection{Low electron fractions across the filament}\label{subsec:xe-result}

%electron fraction morphology
The electron fraction, $x(e)$, shows a variation mostly confined within two orders of magnitude throughout Cloud H (Fig.~\ref{fig:results}, panel d). The morphology of its distribution closely follows that of $R_\mathrm{D}$ and $f_\mathrm{D}$: those regions where $R_\mathrm{D}\gtrsim0.1$ and $f_\mathrm{D}\gtrsim7$ show electron fractions $x(e)\lesssim10^{-8}$, even down to $x(e)\lesssim10^{-9}$ when the depletion factors reach $\sim10$. A clear example is MM7 where most values are within $x(e)\sim10^{-9}-10^{-8}$, which agrees well with the number density reaching its highest values \citep[i.e. $n\gtrsim2\times10^4$~cm$^{-3}$;][]{mckee1989}. Temperature seems to play a secondary role in the variation of the electron fraction behind $R_\mathrm{D}$ and $f_\mathrm{D}$: its values $\sim20$~K towards MM7 have only a marginal influence over $x(e)$, while values up to $\sim10^{-7}$ are reported in the colder ($\sim10-15$~K), but still dense ($n\sim10^4$~cm$^{-3}$) northern and southern parts of the filament.

%the values of x(e) and its comparison to previous results
The wide variety of conditions covered in Cloud H makes $x(e)$ vary within $(0.002-8)\times10^{-7}$ around its median of $1\times10^{-8}$. These electron fractions are lower, or marginally consistent on average, than those estimated using analytical methods in cold clouds \citep[$\sim1-3\times10^{-6}$;][]{dalgarno1984}, low-mass cores \citep[$\sim10^{-8}-10^{-6}$;][]{caselli1998}, sub-parsec molecular cloud cores \citep[$\sim0.5-2\times10^{-8}$;][]{anderson1999}, and sub-parsec filaments \citep[$\sim0.3-30\times10^{-7}$;][]{pineda2024a}. Chemical modelling of the emission from various molecular tracers, both in steady state \citep{schilke1991,deboisanger1996,williams1998} and with a time dependence \citep{shalabiea1995}, instead returns electron fractions in a better agreement with our results, with averages of $\sim10^{-8}$ and lower to upper limits of $\sim2\times10^{-9}$ and $\sim2\times10^{-7}$, respectively.

%mass regime and external effects influencing x(e)
The large variation in values may come from the different mass regimes of the sources and their degree of irradiation. Other works relying on the chemical modelling of tracers in individual sources highlight these influences: low-mass pre-stellar cores exhibit electron fractions within $\sim10^{-9}-10^{-8}$ \citep{caselli2002,maret2007}; high-mass cores hosting or with neighbouring protostars exhibit electron fractions within $\sim0.5-1.5\times10^{-7}$ \citep{bergin1999}. The majority of Cloud H shows $x(e)\lesssim10^{-8}$, consistent with the low-mass pre-stellar cores and the almost absence of point sources along the filament spine; nevertheless, several pixels reach electron fractions $x(e)\gtrsim0.5\times10^{-7}$, suggesting that either some of these neighbouring point sources and protostars or the magnetic field morphology may influence the ion budget.

\subsection{Cosmic-ray ionisation rate in Cloud H}\label{subsec:CRIRcloudH}

%intro: the variation of zeta
An ionisation rate of $\zeta_2\sim10^{-17}$~s$^{-1}$ is still often considered the standard value for CRs propagating throughout the ISM \citep[see discussion in][]{oka2019-review}. However, in the last 30 years, multiple studies reported systematic variations of $\zeta_2$ in several objects. \citet{caselli1998} and \citet{redaelli2025} determined ionisation rates within $\sim10^{-18}-10^{-16}$~s$^{-1}$ in samples of low-mass cores, skewed more towards values of $\sim10^{-17}-10^{-16}$~s$^{-1}$. Studies of individual pre-stellar cores, high-mass envelopes and high-mass star-forming regions returned a similar order of magnitude \citep{maret2007,ivlev2019,sabatini2020,redaelli2021,sabatini2023}. Finally, values above $\sim10^{-16}$~s$^{-1}$ \citep{podio2014}, and up to $10^{-14}$~s$^{-1}$ \citep{ceccarelli2014,fontani2017,cabedo2023}, were reported only in regions with sources capable of locally accelerating CRs \citep[i.e. protostars and outflows;][]{padovani2015,padovani2016}. Despite the large scatter in estimates, these works demonstrated that local conditions in the region can influence the standard ionisation rate from Galactic CRs.

%the variety of CRIRs throughout Cloud H
The same wide range of ionisation conditions applies within Cloud H. The CRIR shows values of $\sim(0.004-4)\times10^{-16}$~s$^{-1}$ around its median of $2.3\times10^{-18}$~s$^{-1}$ throughout the cloud (see Fig.~\ref{fig:results}, panel d). This variation of three orders of magnitude in $\zeta_2$ mirrors that of $x(e)$: the central region comprising the MM7 clump shows the lowest degree of ionisation with values below $\lesssim10^{-18}$~s$^{-1}$, even lower than those reported in other massive clumps; the regions bright in N$_2$D$^+$ (i.e. $I(\mathrm{N_2D^+})\gtrsim1.5$~K~km~s$^{-1}$; see Fig.~\ref{fig:NOEMAmaps}, central panel), besides MM7, exhibit values a factor of a few above $\gtrsim10^{-18}$~s$^{-1}$, closer to those of massive clumps and low-mass cores; the northern and southern part of the filament, where $x(e)\sim10^{-7}$, reach instead values up and above $\sim10^{-17}$~s$^{-1}$: the southern portion is hosts a few outflows powered by the point sources and protostars, as hinted by the broad SiO lines detected in those sub-regions \citep{jimenez-serra2010,liu2026-alma-sio}; the northern portion may be affected by a change in magnetic field morphology and strength instead \citep{liu2018}. Our results in Cloud H demonstrate that parsec-scale filaments are optimal regions to study a variety of different ionisation conditions.

%comparison with parsec-scale filaments
Variations of $\zeta_2$ within a single region were recently studied in other sub-parsec- and parsec-scale filaments. \citet{pineda2024a} determined a median value for the CRIR of $\zeta_2\sim3\times10^{-17}$~s$^{-1}$ in the low-mass NGC~1333 SW filament in Perseus \citep[see e.g.][]{walawender2008} with rates growing to $\sim10^{-16}$~s$^{-1}$ towards protostars in the fields. \citet{socci2024a} reported analogous results towards the massive filament that hosts the Orion molecular clouds 2 and 3 \citep[OMC-2/-3; see e.g.][]{peterson2008}. Despite the common enhancement in $\zeta_2$ seen towards the local sources across the three regions, NGC~1333 SW and OMC-2/-3 still show ionisation rates an order of magnitude higher than those in Cloud H on average. This discrepancy suggests that either the ambient flux of CRs in the former two clouds is constantly supplied (e.g. from the nearby O/B stars) or the flux of CRs in the latter cloud is overall attenuated.

%comparison with GMFs and IRDCs
Ionisation rates in closer agreement to those of Cloud H were recently reported in other IRDCs and giant molecular filaments (GMFs). \citet{entekhabi2022} explored the emission of several molecular tracers in relation to the physical properties of Cloud C. In several positions across the cloud, the models run by the authors reproduced the emission of the tracers for densities $n(\mathrm{H_2})\sim10^4-10^5$~cm$^{-3}$, depletion factors $f_\mathrm{D}\sim3-10$, and ionisation rates of $\zeta_2\sim2-5\times10^{-18}$~s$^{-1}$. \citet{clarke2024} reported similarly low ionisation rates ($\sim2\times10^{-18}$~s$^{-1}$) in the northern part of the GMF G214.5-1.8 for depletion factors $f_\mathrm{D}\sim5$ and densities $n(\mathrm{H_2})\lesssim10^4$~cm$^{-3}$. The lack of O/B stars within a few parsecs, the low number of local young sources, and the similar physical properties ($n$, $f_\mathrm{D}$) shared by these two regions and Cloud H may explain their overall low ionisation rates ($\lesssim10^{-17}$~s$^{-1}$) compared to the aforementioned parsec-scale filaments.

\begin{figure}
    \centering
    \includegraphics[width=0.99\linewidth]{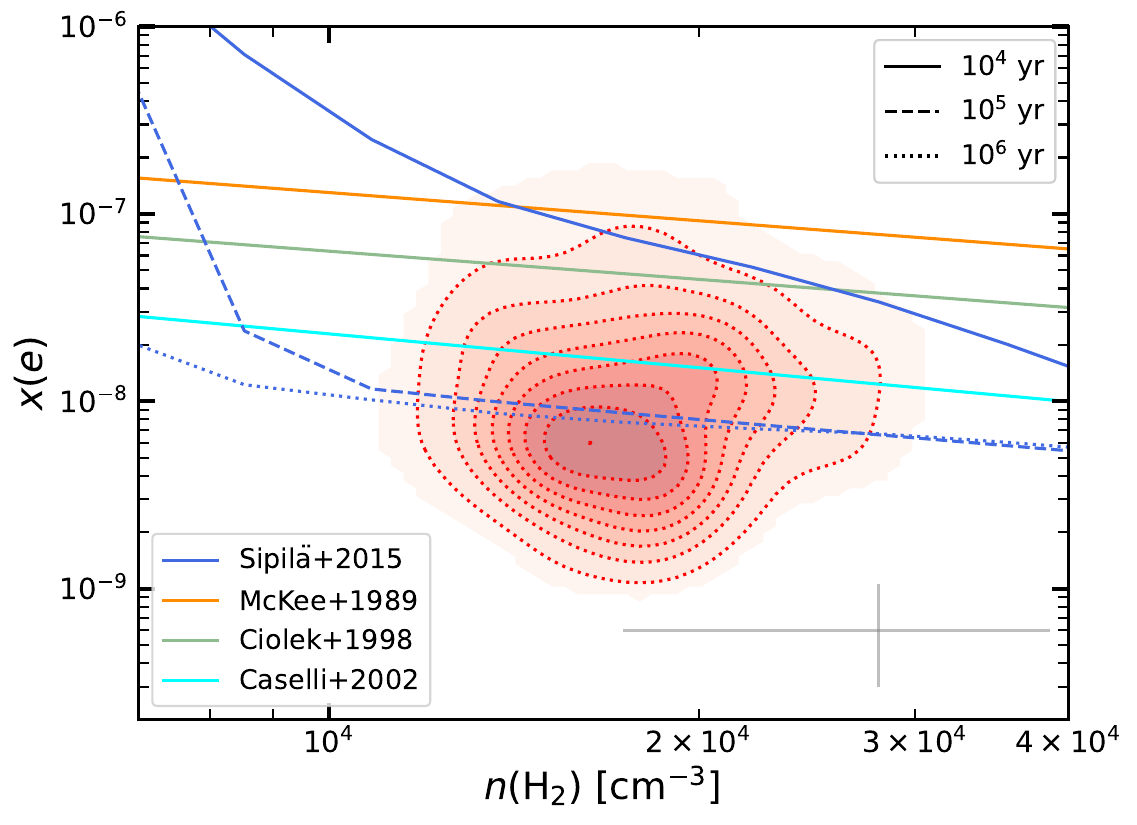}
    \caption{KDE distribution of $x(e)$ with $n(\mathrm{H_2})$ in Cloud H (red shaded area). The lines represent the different relations connecting the two parameters as determined by \citet{mckee1989} (orange), \citet{ciolek1998} (green), and \citep{caselli2002} (cyan). See Sect.~\ref{subsec:dynamic} for the individual relations. The blue lines represent the results from \texttt{pyRate} for different cloud ages \citep{sipila2015} (see text for the setup). The grey cross represents the typical uncertainty on an individual value of $x(e)$.}
    \label{fig:xe_vs_nh2}
\end{figure}

\subsection{Density dependence of the electron fraction}\label{subsec:dynamic}

%ambipolar diffusion
Cosmic rays dominate the ionisation of the gas in clouds with densities above a few visual extinctions \citep{hollenbach1999}. Therefore, the value of the electron fraction is directly driven by $\zeta_2$ in this gas regime, which is also the framework for this study. From this premise and a few assumptions, \citet{mckee1989} demonstrated that the electron fraction and the gas density can be linked via a simple analytical relation:
\begin{equation}\label{eq:mckee}
    x(e)\sim1.3\times10^{-5}n(\mathrm{H_2})^{-0.5}.
\end{equation}
Later studies based on chemical and physical models \citep{bergin1997,caselli1998,caselli2002} demonstrated that additional effects, such as the freeze-out of metals, alter both the pre-factor and the scaling of Eq.~\ref{eq:mckee}. Nevertheless, all relations agree on a net decrease of $x(e)$ for an increase in $n(\mathrm{H_2})$ due to the progressive recombination of ions.

%x(e) vs n(H2) in Cloud H
This anti-correlation between $x(e)$ and $n(\mathrm{H_2})$ applies in Cloud H as well. We explored this dependence through the KDE distribution of the two parameters taken from Fig.~\ref{fig:results}, panels (a) and (c). Figure~\ref{fig:xe_vs_nh2} shows the KDE distribution along with the scaling relations for $x(e)$ from \citet[][orange line; see Eq.~\ref{eq:mckee}]{mckee1989}, \citet[][green line; i.e. $2\times10^{-8}(n(\mathrm{H_2})/10^5~\mathrm{cm}^{-3})^{-0.5}$]{ciolek1998} and \citet[][cyan line; i.e. $5.75\times10^{-6}n(\mathrm{H_2})^{-0.6}$]{caselli2002}. The electron fractions throughout Cloud H show a clear dependence on the number density. When compared to the scaling relations, patterns start to emerge: the relations from \citet{mckee1989} and \citet{ciolek1998} are consistent only with ionisation fractions above $\gtrsim5\times10^{-8}$ for the density regime of Cloud H. These ionisation fractions correspond to those sub-regions in the filament potentially affected by local sources of radiation (see Sect.~\ref{subsec:xe-result}), thus suggesting the two relations to be more suited for active regions in the cloud. The peak of the distribution is instead consistent, within errors, with the relation from \citet{caselli2002}. The latter was determined in the pre-stellar core L1544 by modelling the emission from several ions and for a range of depletion factors within $f_\mathrm{D}\sim3-9$. Our results in Cloud H confirm that this relation is best suited to describe the ionisation budget in regions affected by high CO depletion, even when considering parsec-scale objects.

%full chemical models
As a final test, we compared the electron fraction obtained throughout Cloud H with that from \texttt{pyRate} (see Sect.~\ref{sec:comparison with models}). The electron fraction derived from the models depends on the input CRIR per proton $\zeta$. We therefore run the physical and chemical models with the same setup, but with an ionisation rate $\zeta = \langle\zeta_2\rangle/2.3\sim1\times10^{-18}$~s$^{-1}$, where the relation is discussed by \citet{glassgold1974} and given the median value of $\zeta_2$ in Cloud H (see Sect.~\ref{subsec:error-bud}). Figure~\ref{fig:xe_vs_nh2} shows the comparison between the $x(e)$ measured in the filament and the electron fractions derived by the chemical model for different times (blue lines). The models computed at times $\gtrsim10^5$~yr agree well with our estimates, while those at times $10^4$~yr depart from the measured values of $x(e)$ for the density regime of Cloud H. This result confirms once more the reliability of our analytical methods in the determination of $x(e)$ and $\zeta_2$ at later stages in the cloud evolution (see discussion in Sect.~\ref{sec:comparison with models}). 

\begin{figure}
    \centering
    \includegraphics[width=0.99\linewidth]{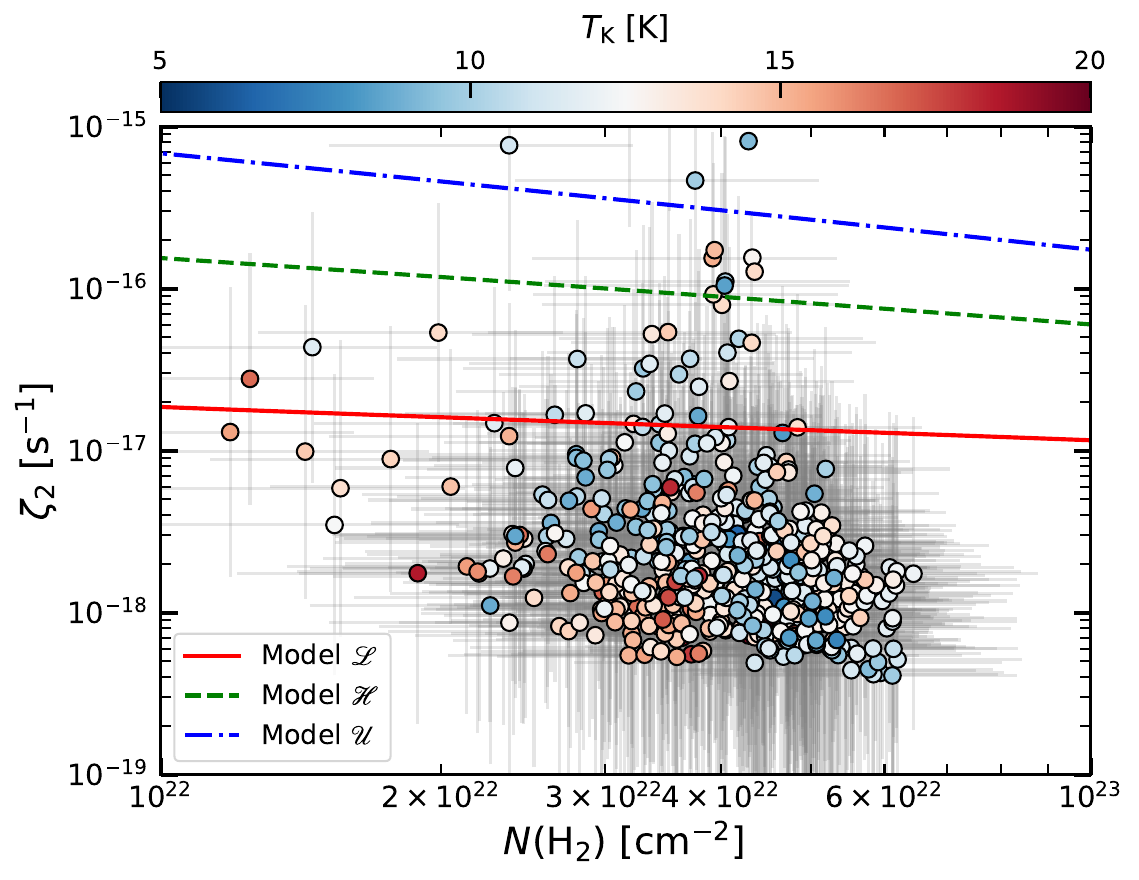}
    \caption{Comparison between $\zeta_2$ and $N(\mathrm{H_2})$, with corresponding errors, in Cloud H. The theoretical predictions for the CR propagation in molecular clouds are displayed as colour-coded lines \citep[blue, green, and red lines; see text for the corresponding model;][]{padovani2024}. The estimates of $\zeta_2$ (Fig.~\ref{fig:results}, panel d) are colour-coded by $T_\mathrm{K}$ (Fig.~\ref{fig:archivaldata}, right panel).}
    \label{fig:padovaniplot}
\end{figure}

\subsection{Cosmic-ray flux attenuation in Cloud H}\label{subsec:padovaniplot}

%models of CR propagation
The ionisation rate of the CRs does not remain constant as the particles penetrate the cloud, but instead decreases with increasing $N(\mathrm{H_2})$. This attenuation was studied by \citet{padovani2009} and comprises theoretical predictions based on different CR propagation and ionisation models sampled in five orders of magnitude in column density. The ionisation is driven by primary CR protons and electrons, plus secondary electrons from primary CRs. Figure~\ref{fig:padovaniplot} displays the CR ionisation models (coloured lines) tailored around the theory of CR propagation in clouds and observational results \citep{padovani2024}: model $\mathscr{L}$ is developed around the CR spectrum observed by the Voyager missions \citep{cummings2016,stone2019}; model $\mathscr{H}$ is developed around the ionisation rates determined in diffuse clouds \citep{indriolo2012,neufeld2017}; model $\mathscr{U}$ is developed around the upper limit of the ionisation rate in those same clouds. Independent estimates of $\zeta_2$ confirm its overall decrease with $N(\mathrm{H_2})$, however, the wide variety of tracers and techniques also causes a broad scatter in the results \citep[see][for a compendium of the estimates]{padovani2024}.

%zeta vs N(H2) in Cloud H
Our work provides new insights on the picture of CR ionisation and flux attenuation, as some recent works did as well \citep[e.g.][]{sabatini2023,pineda2024a,luo2024}: first, it allows for a drastic increase in statistics; then, it probes the density regime within $\sim10^{22}-10^{23}$~cm$^{-2}$, poorly sampled before \citep[see Fig.~C.1 in][]{padovani2022}; finally, it expands on the potential variation of $\zeta_2$ due to local influences within a region. We therefore include in Fig.~\ref{fig:padovaniplot} our estimates of $\zeta_2$ (Fig.~\ref{fig:results}, panel d) with the corresponding values of $N(\mathrm{H_2})$ (Fig.~\ref{fig:archivaldata}, left panel). The overall distribution shows several notable features: first, $\zeta_2$ shows no clear correlation with the temperature besides for the highest values of the latter ($\sim15-20$~K) whose corresponding fields cluster on the lower-left side of the parameter space; then, a functional dependence exists between $N(\mathrm{H_2})$ and $\zeta_2$ across a factor of $\sim3$ in the former (within $\sim2-6\times10^{22}$~cm$^{-2}$) for the coldest fields in our sample ($T_\mathrm{K}\lesssim13$~K) which show ionisation rates below $\lesssim3\times10^{-17}$~s$^{-1}$; finally, outliers to the bulk of the distribution exist for high ionisation rates ($\gtrsim10^{-16}$~s$^{-1}$) with no clear correlation to the column density. None of the predicted CR attenuation models can fully reproduce the distribution of ionisation rates in Cloud H, however a comparison between the two may reveal the nature of these features. 

%comparison with the models: high-temperature fields
Most of the high-temperature fields are neighbouring the point sources and protostars (see Fig.~\ref{fig:archivaldata}, right panel). The latter could locally accelerate CRs on the protostellar surface, thus increasing $\zeta_2$ \citep{padovani2015}. However, these fields show values of $\zeta_2$ up to $\sim10^{-17}$~s$^{-1}$ at best, but overall below model $\mathscr{L}$ for column densities within $\sim2-4\times10^{22}$~cm$^{-2}$. These low values are likely biassed due to their temperatures ($T_\mathrm{K}\sim15-20$~K) being close to the evaporation temperature of CO from the ice coatings of dust grains \citep[e.g.][]{draine2011}. The release of CO in the gas phase reduces its degree of depletion and promotes the destruction of our main tracers, N$_2$H$^+$ and N$_2$D$^+$, potentially causing poorer estimates of both $f_\mathrm{D}$ and $R_\mathrm{D}$.

%comparison with the models: bulk of the distribution - general
Nevertheless, the colder fields, which are the majority, also show anomalous values. Recent re-evaluations of $\zeta_2$ in diffuse clouds \citep{neufeld2024,obolentseva2024} favour the model $\mathscr{L}$ as the more appropriate to describe the typical ionisation rate across the ISM. The colder fields in Cloud H are only in marginal agreement with this model, within errors, given their values lower by a factor of 3 to 9, on average. The deviation of $\zeta_2$ from model $\mathscr{L}$ in Cloud H may originate from different factors: the ionisation rate may be reduced by a combination of large-scale freeze-out of CO at low number densities (i.e. $n(\mathrm{H_2})\sim2000$~cm$^{-3}$) and its location in the Galaxy \citep[as in G214.5-1.8;][]{clarke2024}; the ionisation rate may be reduced by a combination of high column densities and magnetic field gradients \citep[as in Cloud C;][]{entekhabi2022}. However, both the number density regime and the position in the Inner Galaxy \citep[where no decrease of the CR flux is expected;][]{yang2016} make the former interpretation not applicable to Cloud H. Within the latter instead, column densities in Cloud H do not appear to drive its low values of $\zeta_2$ according to the CR propagation models, as shown in Fig.~\ref{fig:padovaniplot}. We thus believe that the magnetic field plays a pivotal role in the attenuation of the CRIR. 

%comparison with the models: bulk of the distribution - magnetic field
\citet{liu2018} studied the magnetic field strength and morphology in Cloud H via observations of the dust polarisation. They reported several properties that can affect the CR propagation and, thus, ionisation: the magnetic field orientation, with respect to the filament spine, changes from mostly parallel towards the North (above the red line in Fig.~\ref{fig:results}, panel d) to mostly perpendicular towards the South (below the red line), with the expected dominant components being poloidal and toroidal, respectively; the magnetic field strength varies from $\sim65~\mu$G throughout the filament to $\sim90-140~\mu$G towards the dense clumps in the region \citep[from North to South, c3 and c4 in][see gold crosses and circles in Fig.~\ref{fig:results}, panel d]{liu2018}. Along converging field lines with a monotonically increasing magnetic field intensity, the focusing and mirroring of the CR pitch angles is balanced. This means that $\zeta_2$ can be calculated as if the field were constant, irrespective of its morphology \citep{padovani2020}. When magnetic field minima and maxima are alternated, magnetic pockets may form and favour mirroring over focusing, thus reducing the CR ionisation by up to an order of magnitude towards the regions with a strong toroidal component \citep{silsbee2018}. We therefore argue that the values of $\zeta_2$ within $(0.5-4)\times10^{-18}$~s$^{-1}$ towards MM7 are a consequence of the uneven magnetic field strength throughout Cloud H, which may form magnetic pockets in the inter-clump filament.

%comparison with the models: the outliers
CRIR values above $\gtrsim10^{-16}$~s$^{-1}$ are potentially affected by local sources instead. These estimates of $\zeta_2$ cross models $\mathscr{H}$ and $\mathscr{U}$, which were adapted to the ionisation rates in diffuse clouds, and lack any correlation with both $N(\mathrm{H_2})$ and $T_\mathrm{K}$. Both arguments support an enhancement of the local CR flux: given the average temperatures of these fields (within $\sim10-15$~K) and their position in the inter-clump parts of the filament (see Fig.~\ref{fig:results}, panel d), these high values of $\zeta_2$ may reflect the presence of CRs mirrored back from the clumps towards the filament, plus their acceleration from the large-scale supernova remnant or local outflows \citep{liu2026-alma-sio}.

\section{Conclusions}\label{sec:conclusions}

We studied the electron fraction and CRIR at parsec scales across the filamentary IRDC G035.39-00.33 (Cloud H) using a newly developed analytical framework. We derived these two parameters using a new set of NOEMA+30m observations for our main suite of tracers, N$_2$H$^+$, N$_2$D$^+$, and DCO$^+$ (1$-$0). These were paired with ancillary observations of C$^{18}$O (1$-$0), plus estimates of the kinetic temperature of the gas, its column and number density of H$_2$ from several facilities, all re-gridded to a common resolution of $3''$ (or $\sim0.04$~pc) and pixel size of $1''$. Our analysis comprised the following steps: we tested our analytical framework against a chemical model to explore its limits of application; then, we estimated the column densities of our molecular tracers; next, we determined the CO depletion factor from the column densities of C$^{18}$O and H$_2$, and the deuterium fraction from the column densities of N$_2$D$^+$ and N$_2$H$^+$. Finally, we combined these estimates to compute the electron fraction and CRIR in Cloud H. The main results of the study are listed below.
\begin{enumerate}
    \item Our analytical framework can accurately reproduce the electron fraction and CRIR from the models within a factor of 3. The formulae to compute $x(e)$ and $\zeta_2$ are most accurate in the number density regime within $\sim10^4-10^5$~cm$^{-3}$ and for cloud ages $\gtrsim10^5$~yr (see Sect.~\ref{sec:comparison with models},~\ref{subsec:assumed-model} and Fig.~\ref{fig:formulavsmodel}).
    \item CO depletion factors above $f_\mathrm{D}\gtrsim3$ are widespread throughout Cloud H with values up to $f_\mathrm{D}\sim12$ towards its densest regions. These depletion factors are higher than those reported in previous studies of Cloud H at lower resolution and show a positive correlation with the number density in the region.
    \item Deuterium fractions above $R_\mathrm{D}\gtrsim0.05$ are similarly widespread in the cloud and its highest values ($\gtrsim0.1$) show a positive correlation with the sites of severe depletion. Deuterium fractions below $R_\mathrm{D}\lesssim0.05$ are found adjacent to evolved compact objects where higher temperatures may reduce the abundance of N$_2$D$^+$ and N$_2$H$^+$ in the gas phase.
    \item The electron fraction varies by two orders of magnitude in Cloud H with values within $\sim10^{-9}-10^{-7}$ and a median of $1\times10^{-8}$. $x(e)$ shows a functional dependence on $n(\mathrm{H_2})$ that is consistent with the scaling relation determined for low-mass cores with depletion factors similar to those of Cloud H and is well-reproduced by the chemical models.
    \item The CRIR in Cloud H varies within $\sim(0.004-4)\times10^{-16}$~s$^{-1}$around its median of $2.3\times10^{-18}$~s$^{-1}$. Most fields experience a lower ionisation compared to the typical value assigned to the ISM (i.e. $10^{-17}$~s$^{-1}$) and below the theoretical predictions for the propagation of CRs in clouds. Their flux attenuation in Cloud H is mildly driven by an increase in $N(\mathrm{H_2})$, while being mostly affected by changes in the magnetic field strength and morphology across the cloud.
    \item Values of $\zeta_2$ above $\gtrsim10^{-16}$~s$^{-1}$ do not show any correlation with $N(\mathrm{H_2})$ nor with $T_\mathrm{K}$. We ascribe these high ionisation rates to the combined effect of CR mirroring, caused by the uneven magnetic field, and acceleration of CRs from a nearby supernova remnant and local outflows in the region.
\end{enumerate}

%closure
Our new analytical framework proved robust in the determination of the electron fraction and CRIR through the wide range of physical properties offered by Cloud H. It challenges other analytical methods \citep{caselli1998,bovino2020,luo2024} thanks to its limited number of assumptions, its accessible tracers, and its wide range of physical conditions probed. Although Cloud H is a perfect benchmark for the method due to its number density regime and cloud age, the results obtained for $x(e)$ and $\zeta_2$ are consistent with independent studies in other regions. Their individual values are, however, more uncertain than the factor of 3 accuracy of the method. While these errors do not undermine the correlations identified in the work, future applications of the method should focus on obtaining better estimates for its most uncertain parameters.

\begin{acknowledgements}
    A.S. is grateful to Stefano Bovino for checking the derivation of the formulae and to Marco Padovani for the discussion on magnetic pockets. A.S. is also indebted to Duo Xu for sharing the number density map of Cloud H. This project has received funding from the European Research Council (ERC) under the European Union’s Horizon 2020 research and innovation programme (Grant agreement No. 851435). J.P., O.S., and P.C. gratefully acknowledge financial support from the Max Planck Society. C-Y.L. acknowledges financial support through the INAF Large Grant `The role of MAGnetic fields in MAssive star formation (MAGMA)'. C-Y.L. also acknowledges support from the Blumberg Astrobiology Grants at the GBT in relation to the reduction and analysis of the GBT data. I.J-.S acknowledges funding from grant PID2022-136814NB-I00 funded by the Spanish Ministry of Science, Innovation and Universities/State Agency of Research MICIU/AEI/ 10.13039/501100011033 and by “ERDF/EU”. J.C.T. acknowledges support from the NSF grant AST-2206450 and SNSA grant SNSA-R 2025-00239. This work is based on observations carried out under project number S21AF001 with the IRAM NOEMA Interferometer and 091-21 with the 30m telescope. IRAM is supported by INSU/CNRS (France), MPG (Germany) and IGN (Spain).
\end{acknowledgements}

\nocite{*}
\bibliographystyle{aa}
\bibliography{new_bib}

%--------------------------------------------

%--------------------------------------------

\begin{appendix}

\section{Chemical network, formulae, and assumed abundances}\label{sec:fullnetwork}

%introduction
The precision of the formulae used to derive $x(e)$ and $\zeta_2$ depends on the number of assumptions we introduce in their determination. At the first level, we included all the reactions involving two among the following reactants: cosmic rays (CR); N$_2$H$^+$, N$_2$D$^+$, and DCO$^+$; the ortho- and para- forms of H$_2$, D$_2$, H$_2^+$, H$_3^+$, H$_2$D$^+$, D$_2$H$^+$, D$_3^+$, plus the meta- form of D$_3^+$; the neutrals CO, N$_2$, HD, O; electron (e$^-$) and negatively-charged grains (g$^-$). The number of reactions that involve two of the latter species accounts for $\sim600$ individual entries. in the following, we will describe the additional selection criteria and assumptions introduced to reduce the number of reactions and obtain the analytical formulae described in Sect.~\ref{sec:analytical formulae}.

\subsection{Rate coefficients and first selection}\label{subsec:rates}

%types of rates
These $\sim$~600 reactions include dissociations, recombinations, and exchange reactions, with correspondingly different rate coefficients. The forms of those relevant in our analysis follow:
\begin{itemize}
    \item dissociative recombination with grains: the rate coefficient takes the form:
    \begin{equation}\label{eq:grain-rate}
        k_g = \eta\sigma Sv_\mathrm{th}J_\mathrm{neg}~,
    \end{equation}
    where: $\eta$ is a tabulated coefficient; $\sigma=\pi a^2$ is the geometric cross section with $a=0.1~\mu\mathrm{m}$ grain radius (see text); $S$ is the sticking coefficient \citep{sipila2015}; $\varv_\mathrm{th}=\sqrt{8k_\mathrm{B}T/\pi\mu m_\mathrm{H}}$ is the thermal speed of the species of molecular weight $\mu$ at temperature $T$; $J_\mathrm{neg}$ is the Coulomb focusing caused by the grain charge \citep[see][for the derivation]{draine1987}.
    \item cosmic-ray-induced dissociation/ionisation: the rate coefficient takes the form:
    \begin{equation}\label{eq:CR-rate}
        k_\mathrm{CR} = k_n\zeta~,
    \end{equation}
    where $k_n$ is a tabulated coefficient and $\zeta$ is the CRIR per proton \citep[i.e. $\zeta=\zeta_2/2.3$;][]{glassgold1974}.
    \item exchange reaction/dissociative recombination with electrons: the rate coefficient takes the typical modified-Arrhenius form:
    \begin{equation}\label{eq:Arr-rate}
        k_\mathrm{Arr} = \alpha\bigg(\frac{T}{300~\mathrm{K}}\bigg)^\beta e^{-\gamma/T}~,
    \end{equation}
    where $\alpha$, $\beta$, and $\gamma$ are tabulated coefficients.
    \item ion-polar system: the rate coefficient takes the form \citep{wakelam2012}:
    \begin{equation}\label{eq:ionpol-rate}
        k_\mathrm{ion} = \alpha\beta\bigg(1 + 0.0967\gamma \bigg(\frac{300}{T}\bigg)^{1/2} + \frac{\gamma^2}{10.526}\frac{300}{T}\bigg)~,
    \end{equation}
    where $\alpha$, $\beta$, and $\gamma$ are tabulated coefficients.
    \item exchange reaction/dissociative recombination with electrons: the rate coefficients for interactions between H$_3^+$ + H$_2$ and isotopologues are tabulated at different temperatures \citep{sipila2017}.
\end{itemize}
The tabulated coefficients are available in the Kinetic Database for Astrochemistry \citep[KIDA\footnote{\url{https://kida.astrochem-tools.org}};][]{wakelam2024}.

%first selection
Equations~\ref{eq:grain-rate},~\ref{eq:CR-rate},~\ref{eq:Arr-rate},~\ref{eq:ionpol-rate}, allow us to compute the rate coefficients, or to interpolate them from those tabulated for the H$_3^+$ + H$_2$ system, at a given temperature. Thus, we determine these rate coefficients at 15~K, the typical temperature found in IRDCs \citep{pillai2006}. Many of the rate coefficients for our sample of $\sim600$ reactions are negligible, therefore, we take an educated guess on which to cut. Assuming a typical abundance for N$_2$H$^+$, with respect to H$_2$, of $x(\mathrm{N_2H^+})\sim10^{-10}$ \citep[e.g.][]{tafalla2002}, a lower limit on the deuterium fraction of $R_\mathrm{lim}\sim0.01$ \citep{crapsi2005}, and a typical rate coefficient for ion-neutral reactions of $k_\mathrm{lim}\sim10^{-10}$~cm$^{-3}$~s$^{-1}$ \citep{herbst1973}, we consider only the rate coefficients $k(15~\mathrm{K})>x(\mathrm{N_2H^+})k_\mathrm{lim}R_\mathrm{lim}\sim10^{-22}$~cm$^{-3}$~s$^{-1}$. This selection reduces the network to 390 reactions with which we determine the analytical formulae.

\subsection{Full derivation of the formulae}\label{subsec:fullderivation}

%derivation: Rd
Let us start with the derivation of the formula for $x(e)$. The first step towards Eq.~\ref{eq:xe} is the determination of $R_\mathrm{D}$ from N$_2$D$^+$ and N$_2$H$^+$, from which Eq.~\ref{eq:main_Rd} follows. We therefore consider the production and destruction pathways for these two species (see Table~\ref{tab:full-network}) and solve their ODEs under the steady-state condition:
\begin{align}\label{eq:steady-state-n2d+}
\begin{split}
    &n({\rm{N_2D^+}})\big[(k_{11}+k_{22})n(e) + k_8n({\rm{CO}})\big] = \\&n({\rm{N_2}})k_{20}\big[n({\rm{H_2D^+}}) + 2n({\rm{D_2H^+}}) + 3n({\rm{D_3^+}})\big]
\end{split}
\end{align}
\begin{align}\label{eq:steady-state-n2h+}
\begin{split}
    &n({\rm{N_2H^+}})\big[(k_{11}+k_{22})n(e) + k_8n({\rm{CO}})\big] = \\&n({\rm{N_2}})k_{20}\big[3n({\rm{H_3^+}}) + 2n({\rm{H_2D^+}}) + n({\rm{D_2H^+}})\big]~,
\end{split}
\end{align}
where we took advantage of the equivalencies between rate coefficients to sum all ortho, para, and meta forms of H$_3^+$ and its isotopologues. Taking the ratio of Eqs.~\ref{eq:steady-state-n2d+},~\ref{eq:steady-state-n2h+}, we obtain the deuterium fraction:
\begin{equation}\label{eq:Rd-full}
    R_{\rm{D}} = \frac{n({\rm{H_2D^+}}) + 2~n({\rm{D_2H^+}}) + 3~n({\rm{D_3^+}})}{3~n({\rm{H_3^+}}) + 2~n({\rm{H_2D^+}}) + n({\rm{D_2H^+}})}~,
\end{equation}
which may be inverted to retrieve Eq.~\ref{eq:main_Rd}.

%derivation: n(DCO+) and x(e)
The next step is to derive the relation between the abundance of DCO$^+$ and that of the H$_3^+$ isotopologues. Taking once more the production and destruction pathways of DCO$^+$ from Table~\ref{tab:full-network}, we can solve the ODE for DCO$^+$ under the steady-state condition obtaining Eq.~\ref{eq:DCO$^+$}. Now, we can substitute Eq.~\ref{eq:DCO$^+$} in Eq.~\ref{eq:main_Rd}, or, even more plainly, in Eq.~\ref{eq:steady-state-n2d+} since both relations share the same sum over the H$_3^+$ isotopologues. The latter substitution leads to the following equation:
\begin{align}\label{eq:n2d+-dco+}
\begin{split}
    &\frac{n({\rm{N_2D^+}})}{n({\rm{N_2}})k_{20}}\big[(k_{11}+k_{12})n(e) + k_8n({\rm{CO}})\big] =\\ &\frac{n(\mathrm{DCO^+})}{n(\mathrm{CO})k_{18}}\big[k_{21}n(e) + k_{22}n(g^-)\big]~.
\end{split}
\end{align}
We can convert the above equation into abundances dividing both sides by $n(\mathrm{H_2})$ and recover Eq.~\ref{eq:xe} by defining $K = \frac{k_{20}}{k_{18}}\frac{x(\mathrm{N_2})}{x(\mathrm{CO})}\frac{x(\mathrm{DCO^+})}{x(\mathrm{N_2D^+})}$ and isolating $x(e)$.

%derivation: zeta from H2+
Let us now continue with the derivation for $\zeta_2$. We first narrowed the interaction between CR particles and the gas to the sole ionisation of H$_2$, thus excluding all other CR-induced reactions. The next step is solving the ODEs for ortho and para H$_2^+$ under the steady-state condition, which gives the relations:
\begin{align}
\begin{split}
    &n(\mathrm{oH_2)}k_1\zeta_2 = n(\mathrm{oH_2^+})\big[3k_2n(\mathrm{H_2)} + \Sigma_xn(x)k_x \big]\\
    &n(\mathrm{pH_2)}k_1\zeta_2 = n(\mathrm{pH_2^+})\big[3k_2n(\mathrm{H_2)} + \Sigma_xn(x)k_x \big]~,
\end{split}
\end{align}
where we already summed ortho and para H$_2$ given their equal rate coefficient and where the sum runs over the other destruction pathways (i.e. $x=\mathrm{CO,~N_2,~e^-~O,...}$). By adding the above relations and taking the root of $n(\mathrm{H_2})$, we obtain the ODE:
\begin{equation}
    k_1\zeta_2 = n(\mathrm{H_2^+})\big[3k_2+\mathcal{O}(10^{-4})\big]~,
\end{equation}
as the abundances of the other species are always $n(\mathrm{X})/n(\mathrm{H_2})\lesssim10^{-4}$. By isolating $n(\mathrm{H_2^+})$, we retrieve Eq.~\ref{eq:h2+}.

%derivation: H3+
The next step is to connect $\zeta_2$ to the abundance of H$_3^+$. We therefore write its ODE in steady-state considering all the production and destruction pathways:
\begin{align}\label{eq:h3+-full}
\begin{split}
    &3k_2n(\mathrm{H_2})n(\mathrm{H_2^+}) +\\ 
    &n(\mathrm{pH_2D^+})\big[\alpha_1n(\mathrm{oH_2})+\beta_1n(\mathrm{pH_2})\big] + \\ &n(\mathrm{oH_2D^+})\big[\alpha_2n(\mathrm{oH_2})+\beta_2n(\mathrm{pH_2})\big] + \\ &n(\mathrm{pD_2H^+})\big[\alpha_3n(\mathrm{oH_2})+\beta_3n(\mathrm{pH_2})\big] + \\ &n(\mathrm{oD_2H^+})\big[\alpha_4n(\mathrm{oH_2})+\beta_4n(\mathrm{pH_2})\big] = \\
    &n(\mathrm{H_3^+})\big[2(k_3+k_4)n(\mathrm{O})+2(k_5+k_6)n(\mathrm{CO})+\\&3k_{20}n(\mathrm{N_2})+an(\mathrm{HD})+bn(\mathrm{D_2})+4k_{13}n(\mathrm{g^-})\big] + \\
    &4n(e)\big[k_9 n(\mathrm{oH_3^+}) + 2k_{10} n(\mathrm{pH_3^+})\big]~,
\end{split}
\end{align}
where $\alpha_x$, $\beta_x$ ($x=1,2,3,4$) are the rate coefficients for the reactions inverse to the deuteration of H$_3^+$ (not shown in Table~\ref{tab:full-network}). To simplify Eq.~\ref{eq:h3+-full} multiple assumptions are required: first, the reactions between the ortho and para forms of H$_2$D$^+$, D$_2$H$^+$, and H$_2$ are assumed to cause a spin-state change more so than the production of H$_3^+$ \citep[based on the results of][]{brunken2014}; second, the steady-state assumption allows us to ignore these spin-state changes as the chemistry is already in equilibrium; third, the electron term shows $2k_{10}/k_9\sim9$ and we can assume $n(\mathrm{pH_3^+})\gtrsim n(\mathrm{oH_3^+})$ in steady state, therefore we neglect the term $k_9n(\mathrm{oH_3^+})$ and take $n(\mathrm{pH_3^+})\sim n(\mathrm{H_2})$. The simplified Eq.~\ref{eq:h3+-full} is read as Eq.~\ref{eq:h3+ simplified} after including $\zeta_2$ from Eq.~\ref{eq:h2+}.

%derivation: Rd limit
The final step is to connect our observables, namely the abundances of N$_2$D$^+$, N$_2$H$^+$, DCO$^+$, and CO, to $\zeta_2$. We can achieve this solution by assuming a low $R_\mathrm{D}$ to simplify Eq.~\ref{eq:main_Rd}:
\begin{equation}\label{eq:rd-simplified}
    n(\mathrm{H_3^+}) \sim\frac{1}{3}\frac{n(\mathrm{H_2D^+}) + 2n(\mathrm{D_2H^+}) + 3n(\mathrm{D_3^+})}{R_\mathrm{D}}~.
\end{equation}
Before continuing with the derivation, let us determine which value of $R_\mathrm{D}$ corresponds to this low limit. Taking the ratio of Eq.~\ref{eq:main_Rd}, which represents the total abundance of $n(\mathrm{H_3^+})$, $n_\mathrm{full}$, and Eq.~\ref{eq:rd-simplified}, which represents its approximate value, $n_\mathrm{app}$, we obtain the following relation:
\begin{equation}\label{eq:approx}
    \frac{n_\mathrm{full}}{n_\mathrm{app}} = 1-2R_\mathrm{D}-R_\mathrm{D}\frac{n(\mathrm{D_2H^+}) + 6n(\mathrm{D_3^+})}{n(\mathrm{H_2D^+}) + 2n(\mathrm{D_2H^+}) + 3n(\mathrm{D_3^+})}~.
\end{equation}
Different density regimes see an H$_3^+$ isotopologue dominating over the other two \citep{flower2006} and Eq.~\ref{eq:approx} provides the corresponding error in using the approximate formula:
\begin{align}
\begin{split}
    \mathrm{H_2D^+}:&\frac{n_\mathrm{full}}{n_\mathrm{app}} \sim 1-2R_\mathrm{D}~,\qquad n(\mathrm{H_2})\sim10^4-10^5~\mathrm{cm^{-3}}~,\\
    \mathrm{D_2H^+}:&\frac{n_\mathrm{full}}{n_\mathrm{app}} \sim 1-\frac{5}{2}R_\mathrm{D}~,\qquad n(\mathrm{H_2})\sim10^5~\mathrm{cm^{-3}}~,\\
    \mathrm{D_3^+}:&\frac{n_\mathrm{full}}{n_\mathrm{app}} \sim 1-4R_\mathrm{D}~,\qquad n(\mathrm{H_2})>10^5~\mathrm{cm^{-3}}~.
\end{split}
\end{align}
Our estimate of $\zeta_2$ will not overshoot the actual value by more than a factor of 3 due to this approximation at any number density for deuterium fractions $R_\mathrm{D}\lesssim0.15$. This limit may increase to values of $R_\mathrm{D}\lesssim0.3$ if $\mathrm{H_2D^+}$ dominates, namely if the number density remains $\lesssim10^5$~cm$^{-3}$.

%derivation: zeta
We recognise in Eq.~\ref{eq:rd-simplified} the same sum over the H$_3^+$ isotopologues used in the ODEs of N$_2$D$^+$ and DCO$^+$. By substituting the latter into Eq.~\ref{eq:rd-simplified}, the abundance of H$_3^+$ reads as Eq.~\ref{eq:h3+-final}, from which the derivation of $\zeta_2$ using Eq.~\ref{eq:h3+ simplified} is straightforward.

\subsection{Assumed atomic and molecular abundances}\label{subsec:abundances-not-probed}

%intro
Equations~\ref{eq:xe} and \ref{eq:zeta} comprise several parameters that we can probe, described in Sect.~\ref{sec:analysis}, but some we cannot probe as well. For the latter, we will assume typical abundances with respect to H$_2$ reported in the literature and discuss their potential depletion in relation to that of CO.

%assumed abundances
HD and D$_2$, have abundances of $3.2\times10^{-5}$ and $4\times10^{-6}$, respectively. Despite the complex deuterium chemistry of H$_2$ and H$_3^+$, the abundances of these species are rather constant for densities up to and above $n\gtrsim10^5$~cm$^{-3}$ \citep{walmsley2004,flower2004}. The abundances of atomic nitrogen (N) and oxygen (O), with respect to H, were reported as $7.5\times10^{-5}$ \citep{meyer1997} and $3.25\times10^{-4}$ \citep{meyer1998} from spectrographic measurements towards several sight lines. Later estimates and effective freeze-out onto dust grains \citep[e.g.][]{caux1999,vastel2000} suggested a change in these values. We adopt abundances with respect to H$_2$ hereafter, starting from N$_2$ at $x(\mathrm{N_2})=7.56\times10^{-5}$ \citep{semenov2010}. For atomic O, the choice is complex, as this species is progressively locked in CO and H$_2$O. CO locks around $x(\mathrm{CO})\sim10^{-4}$ of this oxygen, while H$_2$O shows variations in abundance within $\sim(0.001-1)\times10^{-4}$ \citep{vandishoeck2021}. The corresponding variation in the abundance of atomic O is $x(\mathrm{O})=(2-4)\times10^{-4}$. We adopt a value of $x(\mathrm{O})=2\times10^{-4}$, although a variation of up to a factor of 2 in $x(\mathrm{O})$ accounts for only a variation of $\sim30\%$ in $\zeta_2$. The abundance of grains available for recombination reactions (i.e. negatively-charged grains) can be computed as follows \citep{mckee1989}:
\begin{equation}\label{eq:grain-abund}
    x(g^-) = 6\times10^{-11}\bigg(\frac{10^{-6}~\mathrm{cm}}{a_\mathrm{min}}\bigg)^{2.5},
\end{equation}
where $a$ is the minimum grain size in the distribution. Assuming a constant grain size of $0.1~\mu$m, as we did to run the models (see Sect.~\ref{sec:comparison with models}), the typical abundance of the negatively-charged grains is $x(g^-)\sim2\times10^{-13}$. We therefore include the grain contribution in Eqs.~\ref{eq:xe} and \ref{eq:zeta}, although its influence is limited compared to the other species for number densities below $\lesssim10^6$~cm$^{-3}$.

%depletion factors
The atomic and molecular abundances of some of these species may be altered by the freeze-out of these species onto dust grains. However, the only depletion factor that we can directly measure is the CO depletion factor ($f_\mathrm{D}$). This leaves the degree of depletion for O ($f_\mathrm{O}$) and N$_2$ ($f_\mathrm{N_2}$) unchecked. The easiest way to treat the latter two factors is to assume that $f_\mathrm{N_2}=1$ and $f_\mathrm{O}=f_\mathrm{D}$. The first assumption is rooted in the slow formation of N$_2$ \citep[e.g.][]{aikawa2005}, for which we consider its abundance unaltered in Cloud H while that of CO is already reduced by a factor $f_\mathrm{D}$ \citep[see discussions in][on the abundance of N$_2$ in two different density regime]{bergin2002,redaelli2019}. The second assumption is rooted in the freeze-out of oxygen onto dust grains under the same conditions and timescales as its main molecular compound, CO \citep{caselli1998}. Despite the abundance of oxygen in the ISM being still debated \citep[see][for a full account on the topic]{vandishoeck2021}, this is the simplest approach for a dense filament such as Cloud H.

\begin{figure}
    \centering
    \includegraphics[width=0.99\linewidth]{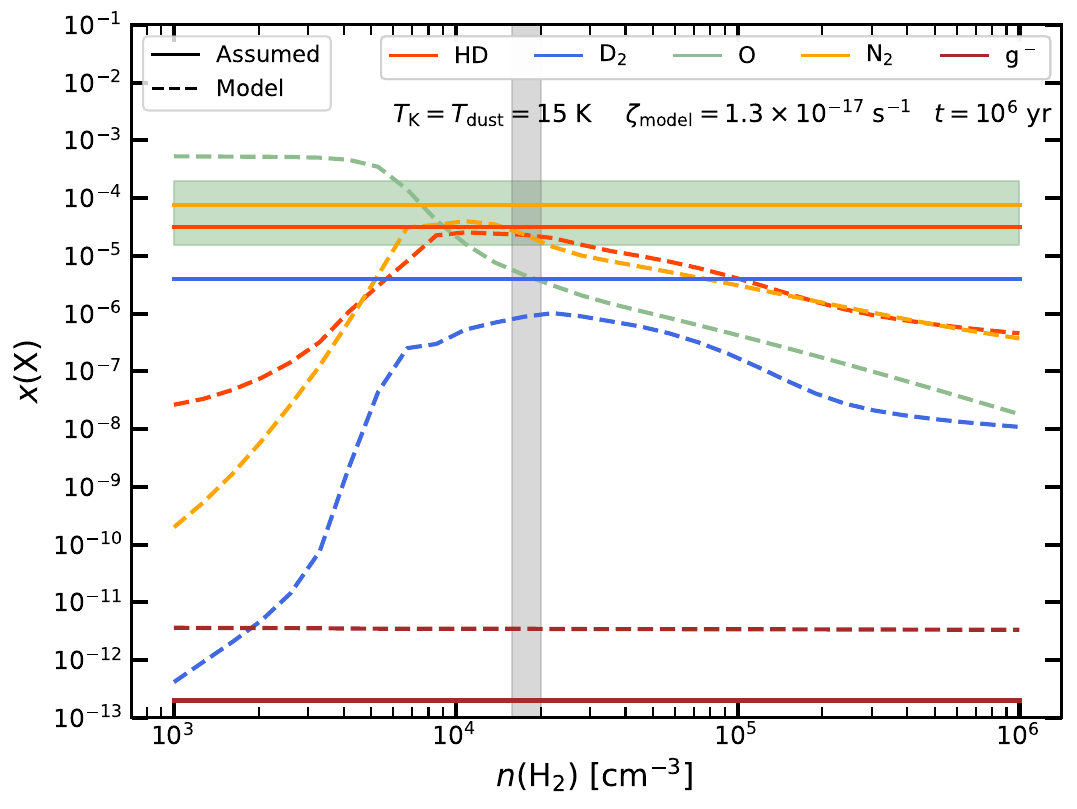}
    \caption{Variation of the abundances of HD, D$_2$, O, N$_2$ and $g^-$ with the number density derived by \texttt{pyRate} (dashed lines) for a time $10^6$~yr and using the same physical model as in Sect.~\ref{sec:comparison with models}. The assumed abundances for the same species used in the analytical calculations (Table~\ref{tab:model params}) are displayed with solid lines instead. The green shaded area represents the variation in the O abundance when applying the CO depletion factor to it (see Sect.~\ref{subsec:abundances-not-probed}). The gray shaded area represents the interquartile range around the median number density in Cloud H.}
    \label{fig:assumedab-modelab}
\end{figure}

\subsection{Cross-check between assumed and modelled abundances}\label{subsec:assumed-model}

%intro: what is in the figure
We finally tested the reliability of our assumed abundances when compared to those derived from \texttt{PyRate}. The outputs considered are those computed at $t=10^6$~yr given the steady-state assumption made when deriving the analytical formulae (see Sect.~\ref{sec:analytical formulae}). Figure~\ref{fig:assumedab-modelab} shows the abundances of HD, D$_2$, O, N$_2$, and $g^-$ as derived from the chemical model (dashed lines) and assumed in our analytic approach (solid lines\footnote{The abundance assumed for O (i.e. $2\times10^4$) has been divided by the range of depletion factors derived in the region, and therefore, it is shown as a green band.}).

%comparison: molecular species and O
The abundances of HD, D$_2$, O and N$_2$ from the model may be overestimated by more than an one order of magnitude compared by those we assumed for values below $\lesssim6\times10^3$~cm$^{-3}$ and above $\lesssim10^5$~cm$^{-3}$. The former regime is dominated by the unattenuated, or marginally attenuated, UV flux, which causes the molecular species to disappear from the gas (HD, D$_2$, N$_2$) while the atomic species remain abundant due to the absence of depletion. The latter regime sees depletion affecting the abundances of N$_2$ and O instead, while those of HD and D$_2$, reduced by the deuteration of ions, are replenished by D and become progressively constant \citep{flower2004}. For number densities within $\sim6\times10^3-10^5$~cm$^{-3}$, the abundances of HD, D$_2$, O and N$_2$ we assumed are in closer agreement with those from the model within a factor of $\lesssim9$. This discrepancy is further reduced to a factor of $\sim3$ when considering the typical number density regime sampled in Cloud H, namely $\langle n(\mathrm{H_2})\rangle=1.8^{+0.2}_{-0.2}\times10^4$~cm$^{-3}$ (as median and interquartile range, IQR). Correcting the abundances downwards by a factor of $\sim9$ does cause a variation in the estimates of $x(e)$ and $\zeta_2$ by less than a factor of $\sim2$ for all species. Although we note that, while varying the abundance of N$_2$ causes only a marginal change in $\zeta_2$, its low abundances reduce the available solutions given the condition in Eq.~\ref{eq:existence-cond}.

%comparison: negatively-charged grains
In contrast to the modelled abundances of HD, D$_2$, O and N$_2$, that of the negatively-changed grains remains constant at $x(g^-)=3.2\times10^{-12}$ across number densities. This value is a factor of $\sim17$ higher than the analytical estimate we performed following Eq.~\ref{eq:grain-abund} \citep{mckee1989}. The order-of-magnitude difference likely arises from the different prescriptions for grain coagulation and distribution considered in the analytical calculation \citep{mathis1977,draine1985} and in the model \citep{taquet2014}. Despite the substantial difference in values, the impact of the grain abundance on $x(e)$ and $\zeta_2$ is marginal: as discussed in Sect.~\ref{sec:comparison with models}, the contribution of grains becomes dominant for densities $n\gtrsim10^6$~cm$^{-3}$, but for the number densities sampled by our analysis the net variation of $x(e)$ and $\zeta_2$ amounts to a few percent when computed with either one of the two grain abundances.

%conclusion: limits of the estimates
Our exploration of the parameter space revealed that our analytical approach is only marginally sensitive to the abundances fed into it. The assumptions on their values cause little variation of the results in a slightly shifted and narrower regime of densities ($\sim6\times10^3-10^5$~cm$^{-3}$) compared to the optimal one claimed for the method (i.e. $10^4-10^6$~cm$^{-3}$; see Sect.~\ref{sec:comparison with models}). These variations are well within its nominal accuracy (a factor of $\sim3$) and that on the individual values of $x(e)$ (a factor of $\sim3$) and $\zeta_2$ (a factor of $\sim9$). However, an educated guess on these abundances is advised to retain the dynamic range in density and, in particular for that of N$_2$, a good number of physical solutions.

\tablefirsthead{\toprule\toprule Reaction & $k(15~\mathrm{K})$ & rate-ID \\ \midrule}
\tablehead{\toprule\toprule Reaction & $k(15~\mathrm{K})$ & rate-ID \\ \midrule}
\tabletail{\midrule}
\tablelasttail{\bottomrule}

\bigskip
\begin{center}
\captionsetup{position=above}
\captionof{table}{Chemical network considered in the derivation of Eqs.~\ref{eq:xe} and \ref{eq:zeta}. From left to right, the columns outline the reaction, its rate at 15~K, the type of the reaction (see text), and the rate coefficient name (rate-ID; as used in the rate equations).}
\begin{supertabular}{lcc}\label{tab:full-network}
%\shrinkheight{-90pt}\noindent
oH$_2$ + CR $\rightarrow$ oH$_2^+$ + e$^-$ & 9.30e-01$\zeta_2$ & $k_1\zeta_2$ \\
pH$_2$ + CR $\rightarrow$ pH$_2^+$ + e$^-$ & 9.30e-01$\zeta_2$ & $k_1\zeta_2$ \\
oH$_2$ + oH$_2^+$ $\rightarrow$ H + oH$_3^+$ & 1.40e-09 & 2$k_2$ \\
oH$_2$ + oH$_2^+$ $\rightarrow$ H + pH$_3^+$ & 7.00e-10 & $k_2$ \\
oH$_2$ + pH$_2^+$ $\rightarrow$ H + oH$_3^+$ & 7.00e-10 & $k_2$ \\
oH$_2$ + pH$_2^+$ $\rightarrow$ H + pH$_3^+$ & 1.40e-09 & 2$k_2$ \\
pH$_2$ + oH$_2^+$ $\rightarrow$ H + oH$_3^+$ & 7.00e-10 & $k_2$ \\
pH$_2$ + oH$_2^+$ $\rightarrow$ H + pH$_3^+$ & 1.40e-09 & 2$k_2$ \\
pH$_2$ + pH$_2^+$ $\rightarrow$ H + pH$_3^+$ & 2.10e-09 & 3$k_2$ \\
O + oH$_3^+$ $\rightarrow$ H + oH$_2$O$^+$ & 4.97e-10 & 2$k_3$ \\
O + pH$_3^+$ $\rightarrow$ H + oH$_2$O$^+$ & 2.48e-10 & $k_3$ \\
O + pH$_3^+$ $\rightarrow$ H + pH$_2$O$^+$ & 2.48e-10 & $k_3$ \\
O + oH$_3^+$ $\rightarrow$ oH$_2$ + OH$^+$ & 1.16e-09 & 2$k_4$ \\
O + pH$_3^+$ $\rightarrow$ oH$_2$ + OH$^+$ & 5.80e-10 & $k_4$ \\
O + pH$_3^+$ $\rightarrow$ pH$_2$ + OH$^+$ & 5.80e-10 & $k_4$ \\
CO + oH$_3^+$ $\rightarrow$ oH$_2$ + HCO$^+$ & 2.31e-09 & 2$k_5$ \\
CO + pH$_3^+$ $\rightarrow$ oH$_2$ + HCO$^+$ & 1.15e-09 & $k_5$ \\
CO + pH$_3^+$ $\rightarrow$ pH$_2$ + HCO$^+$ & 1.15e-09 & $k_5$ \\
CO + oH$_3^+$ $\rightarrow$ oH$_2$ + HOC$^+$ & 1.35e-10 & 2$k_6$ \\
CO + pH$_3^+$ $\rightarrow$ oH$_2$ + HOC$^+$ & 6.77e-11 & $k_6$ \\
CO + pH$_3^+$ $\rightarrow$ pH$_2$ + HOC$^+$ & 6.77e-11 & $k_6$ \\
N$_2$ + oH$_3^+$ $\rightarrow$ oH$_2$ + N$_2$H$^+$ & 1.70e-09 & 3$k_{20}$ \\
N$_2$ + pH$_3^+$ $\rightarrow$ oH$_2$ + N$_2$H$^+$ & 8.50e-10 & 3/2$k_{20}$ \\
N$_2$ + pH$_3^+$ $\rightarrow$ pH$_2$ + N$_2$H$^+$ & 8.50e-10 & 3/2$k_{20}$ \\
CO + N$_2$H$^+$ $\rightarrow$ N$_2$ + HCO$^+$ & 1.07e-09 & $k_8$ \\
e$^-$ + oH$_3^+$ $\rightarrow$ H + H + H & 3.88e-08 & 3$k_9$ \\
e$^-$ + pH$_3^+$ $\rightarrow$ H + H + H & 3.44e-07 & 6$k_{10}$ \\
e$^-$ + oH$_3^+$ $\rightarrow$ oH$_2$ + H & 1.29e-08 & $k_9$ \\
e$^-$ + pH$_3^+$ $\rightarrow$ oH$_2$ + H & 5.73e-08 & $k_{10}$ \\
e$^-$ + pH$_3^+$ $\rightarrow$ pH$_2$ + H & 5.73e-08 & $k_{10}$ \\
N$_2$H$^+$ + e$^-$ $\rightarrow$ N + NH & 1.61e-07 & $k_{11}$ \\
N$_2$H$^+$ + e$^-$ $\rightarrow$ H + N$_2$ & 3.06e-06 & $k_{12}$ \\
g- + oH$_3^+$ $\rightarrow$ g + H + oH$_2$ & 1.22e-04 & 2$k_{13}$ \\
g- + pH$_3^+$ $\rightarrow$ g + H + oH$_2$ & 6.11e-05 & $k_{13}$ \\
g- + pH$_3^+$ $\rightarrow$ g + H + pH$_2$ & 6.11e-05 & $k_{13}$ \\
g- + oH$_3^+$ $\rightarrow$ g + H + H + H & 1.22e-04 & 2$k_{13}$ \\
g- + pH$_3^+$ $\rightarrow$ g + H + H + H & 1.22e-04 & 2$k_{13}$ \\
CO + oH$_2$D$^+$ $\rightarrow$ oH$_2$ + DCO$^+$ & 7.70e-10 & $k_{18}$ \\
CO + pH$_2$D$^+$ $\rightarrow$ pH$_2$ + DCO$^+$ & 7.70e-10 & $k_{18}$ \\
CO + oD$_2$H$^+$ $\rightarrow$ HD + DCO$^+$ & 1.54e-09 & 2$k_{18}$ \\
CO + pD$_2$H$^+$ $\rightarrow$ HD + DCO$^+$ & 1.54e-09 & 2$k_{18}$ \\
CO + mD$_3^+$ $\rightarrow$ oD$_2$ + DCO$^+$ & 2.31e-09 & 3$k_{18}$ \\
CO + pD$_3^+$ $\rightarrow$ pD$_2$ + DCO$^+$ & 2.31e-09 & 3$k_{18}$ \\
CO + oD$_3^+$ $\rightarrow$ oD$_2$ + DCO$^+$ & 1.15e-09 & 3/2$k_{18}$ \\
CO + oD$_3^+$ $\rightarrow$ pD$_2$ + DCO$^+$ & 1.15e-09 & 3/2$k_{18}$ \\
N$_2$ + oH$_2$D$^+$ $\rightarrow$ oH$_2$ + N$_2$D$^+$ & 5.67e-10 & $k_{20}$ \\
N$_2$ + pH$_2$D$^+$ $\rightarrow$ pH$_2$ + N$_2$D$^+$ & 5.67e-10 & $k_{20}$ \\
N$_2$ + oH$_2$D$^+$ $\rightarrow$ HD + N$_2$H$^+$ & 1.13e-09 & 2$k_{20}$ \\
N$_2$ + pH$_2$D$^+$ $\rightarrow$ HD + N$_2$H$^+$ & 1.13e-09 & 2$k_{20}$ \\
N$_2$ + oD$_2$H$^+$ $\rightarrow$ HD + N$_2$D$^+$ & 1.13e-09 & 2$k_{20}$ \\
N$_2$ + pD$_2$H$^+$ $\rightarrow$ HD + N$_2$D$^+$ & 1.13e-09 & 2$k_{20}$ \\
N$_2$ + oD$_2$H$^+$ $\rightarrow$ oD$_2$ + N$_2$H$^+$ & 5.67e-10 & $k_{20}$ \\
N$_2$ + pD$_2$H$^+$ $\rightarrow$ pD$_2$ + N$_2$H$^+$ & 5.67e-10 & $k_{20}$ \\
N$_2$ + mD$_3^+$ $\rightarrow$ oD$_2$ + N$_2$D$^+$ & 1.70e-09 & 3$k_{20}$ \\
N$_2$ + pD$_3^+$ $\rightarrow$ pD$_2$ + N$_2$D$^+$ & 1.70e-09 & 3$k_{20}$ \\
N$_2$ + oD$_3^+$ $\rightarrow$ oD$_2$ + N$_2$D$^+$ & 8.50e-10 & 3/2$k_{20}$ \\
N$_2$ + oD$_3^+$ $\rightarrow$ pD$_2$ + N$_2$D$^+$ & 8.50e-10 & 3/2$k_{20}$ \\
CO + N$_2$D$^+$ $\rightarrow$ N$_2$ + DCO$^+$ & 1.07e-09 & $k_8$ \\
DCO$^+$ + e$^-$ $\rightarrow$ D + CO & 2.21e-06 & $k_{21}$ \\
N$_2$D$^+$ + e$^-$ $\rightarrow$ N + ND & 1.61e-07 & $k_{11}$ \\
N$_2$D$^+$ + e$^-$ $\rightarrow$ D + N$_2$ & 3.06e-06 & $k_{12}$ \\ 
g$^-$ + DCO$^+$ $\rightarrow$ g + CO + D & 7.76e-05 & $k_{22}$ \\
\hline
\multicolumn{3}{c}{$a=\Sigma_ik_i$, with $i=1,2,3,4$}\\
pH$_3^+$ + HD $\rightarrow$ pH$_2$D$^+$ + pH$_2$ & 3.17e-10 & - \\
pH$_3^+$ + HD $\rightarrow$ pH$_2$D$^+$ + oH$_2$ & 4.54e-10 & - \\
pH$_3^+$ + HD $\rightarrow$ oH$_2$D$^+$ + pH$_2$ & 6.45e-10 & - \\
pH$_3^+$ + HD $\rightarrow$ oH$_2$D$^+$ + oH$_2$ & 1.03e-10 & - \\
\hline
\multicolumn{3}{c}{$b=\Sigma_ik_i$, with $i=1,2,3,4$}\\
pH$_3^+$ + pD$_2$ $\rightarrow$ pH$_2$D$^+$ + HD & 3.58e-10 & - \\
pH$_3^+$ + pD$_2$ $\rightarrow$ oH$_2$D$^+$ + HD & 5.11e-10 & - \\
pH$_3^+$ + pD$_2$ $\rightarrow$ pD$_2$H$^+$ + pH$_2$ & 3.04e-10 & - \\
pH$_3^+$ + pD$_2$ $\rightarrow$ pD$_2$H$^+$ + oH$_2$ & 3.89e-10 & - \\
\hline
\shrinkheight{-160pt}
\multicolumn{3}{c}{$b=\Sigma_ik_i$, with $i=1,2,3,4$}\\
pH$_3^+$ + oD$_2$ $\rightarrow$ pH$_2$D$^+$ + HD & 2.95e-10 & - \\
pH$_3^+$ + oD$_2$ $\rightarrow$ oH$_2$D$^+$ + HD & 2.38e-10 & - \\
pH$_3^+$ + oD$_2$ $\rightarrow$ oD$_2$H$^+$ + pH$_2$ & 4.96e-10 & - \\
pH$_3^+$ + oD$_2$ $\rightarrow$ oD$_2$H$^+$ + oH$_2$ & 5.38e-10 & - \\
\hline
\multicolumn{3}{c}{$a=\Sigma_ik_i$, with $i=1,2,3$}\\
oH$_3^+$ + HD $\rightarrow$ pH$_2$D$^+$ + oH$_2$ & 1.74e-10 & - \\
oH$_3^+$ + HD $\rightarrow$ oH$_2$D$^+$ + pH$_2$ & 2.29e-10 & - \\
oH$_3^+$ + HD $\rightarrow$ oH$_2$D$^+$ + oH$_2$ & 1.09e-09 & - \\
\hline
\multicolumn{3}{c}{$b=\Sigma_ik$, with $i=1,2$}\\
oH$_3^+$ + pD$_2$ $\rightarrow$ oH$_2$D$^+$ + HD & 7.96e-10 & - \\
oH$_3^+$ + pD$_2$ $\rightarrow$ pD$_2$H$^+$ + oH$_2$ & 7.56e-10 & - \\
\hline
\multicolumn{3}{c}{$b=\Sigma_ik_i$, with $i=1,2$}\\
oH$_3^+$ + oD$_2$ $\rightarrow$ oH$_2$D$^+$ + HD & 4.77e-10 & - \\
oH$_3^+$ + oD$_2$ $\rightarrow$ oD$_2$H$^+$ + oH$_2$ & 1.09e-09 & - \\
\end{supertabular}
\end{center}

%\vspace{-1ex}
\section{Additional maps and tests}\label{sec:additional maps}

%intro
We report in this Appendix the description of the ancillary data used in the paper and additional tests performed on a few estimates: we first outline the properties of $T_\mathrm{K}$ and $N(\mathrm{H_2})$ towards Cloud H and presented in Sect.~\ref{subsec:ancillary}, we then discuss the CR flux attenuation, presented in Sect.~\ref{subsec:padovaniplot}, by considering the $\zeta_2/n(\mathrm{H_2})-N(\mathrm{H_2})$ parameter space instead.

\begin{figure*}
    \centering
    \includegraphics[width=0.99\linewidth]{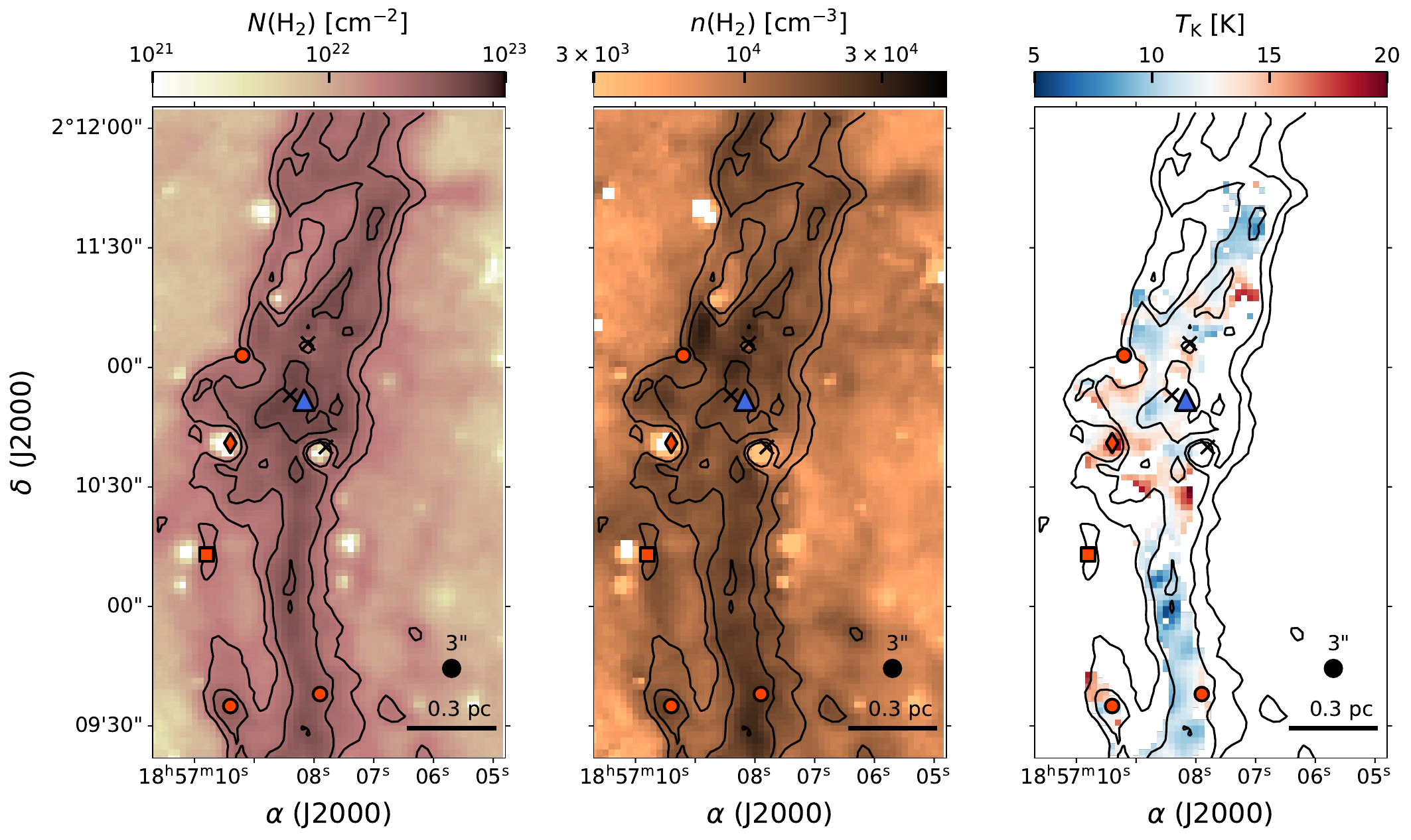}
    \caption{Ancillary observations of Cloud H. \textit{left panel}: column density map, $N(\mathrm{H_2})$, derived from the NIR+MIR extinction map of \citet{kainulainen2013-maps}. \textit{central panel}: number density map, $n(\mathrm{H_2})$, derived from the total column density using the denoising diffusion probabilistic model \citep[DDPM;][]{xu2023}. \textit{right panel}: kinetic temperature map obtained from VLA+GBT observations of the NH$_3$ (1,1), (2,2) inversion transitions \citep{sokolov2019}. In all panels, the contours correspond to H$_2$ column densities of [$2\times10^{22}$, $3\times10^{22}$, $5\times10^{22}$]~cm$^{-2}$ and the symbols are the same as in Fig.~\ref{fig:NOEMAmaps}.}
    \label{fig:archivaldata}
\end{figure*}

\subsection{Column density, number density, and kinetic temperature maps}\label{subsec:additional-maps}

%morphology of N(H2)
The cloud appears in $N(\mathrm{H_2})$ as a narrow and spatially-coherent filament ($\sim0.2-0.3$~pc) with a length of $\sim2$~parsecs (Fig.~\ref{fig:archivaldata}, left panel). This monolithic structure, encompassing column densities within $\sim1-6\times10^{22}$~cm$^{-2}$, shows breaks at two opposite regimes: for $N(\mathrm{H_2})\sim10^{21}$~cm$^{-2}$, the point sources with bright emission at $24~\mu\mathrm{m}$ \citep{gutermuth2015} cannot be recovered by the NIR+MIR method producing these artificially low-density sub-regions (see discussion in Sect.~\ref{subsec:ancillary}); for $N(\mathrm{H_2})\gtrsim5\times10^{22}$~cm$^{-2}$, several dense sub-structures emerge from the filament spine closely following those detected in N$_2$D$^+$, and among which we recognise MM7 as the prominent central one \citep{rathborne2006}.

%the number density map
The number density map (Fig.~\ref{fig:archivaldata}, central panel) closely follows the column density one (Fig.~\ref{fig:archivaldata}, left panel), retaining its main features. The number densities span around an order of magnitude throughout Cloud H, varying within $\sim0.4-4\times10^4$~cm$^{-3}$. The median and interquartile range (IQR) for the number density are $\langle n(\mathrm{H_2})\rangle\sim1.0^{+0.4}_{-0.3}\times10^4$~cm$^{-3}$, which grows and narrows to $\langle n(\mathrm{H_2})\rangle\sim1.8^{+0.2}_{-0.2}\times10^4$~cm$^{-3}$ when considering only the filament spine (i.e. for the lowest contour in Fig.~\ref{fig:archivaldata}, left panel)\footnote{These values are obtained once the artificially low number densities nearby the point sources are masked, namely considering only those pixels with $N(\mathrm{H_2})\geq2\times10^{21}$~cm$^{-2}$.}. These number densities are consistent with those reported by \citet{jimenez-serra2014} (i.e. within $\sim0.4-1.3\times10^4$~cm$^{-3}$) from an LVG analysis of multiple C$^{18}$O transitions. Our estimates are around an order of magnitude below independent number densities measured in another parsec-scale filament at similar spatial resolution \citep[$\sim0.06$~pc;][]{socci2024a} instead. The difference likely arises from the higher column densities sampled in this latter study (i.e. $\sim10^{23}$~cm$^{-3}$). Finally, number densities of $\sim10^4$~cm$^{-3}$ confirm Cloud H as a suitable target to achieve reliable estimates of $x(e)$ and $\zeta_2$ for cloud ages above $>10^4$~yr (see Fig.~\ref{fig:formulavsmodel}).

%the temperature in Cloud H
The kinetic temperature probed by NH$_3$ correlates with the densest spine of Cloud H (Fig.~\ref{fig:archivaldata}, right panel). This arrangement is expected given the spatial distribution of the N$_2$H$^+$ emission (see Fig.~\ref{fig:NOEMAmaps}) and the precursor-derivative connection between the two species \citep{aikawa2005}\footnote{We note, however, that the co-spatial existence of NH$_3$ and N$_2$H$^+$ in the plane of the sky may not be mirrored along the line of sight since their linewidths diverge by $\sim30\%$ on average \citep{sokolov2019}}. Although associated with the densest parts of the cloud, the temperature experiences significant variations across the filament showing values within $\sim7-25$~K. The coldest positions in Cloud H ($T_\mathrm{K}\lesssim10$~K) coincide with gas structures having densities of $N(\mathrm{H_2})\gtrsim5\times10^{22}$~cm$^{-2}$ in most cases. Outside of these sub-regions, $T_\mathrm{K}$ rises to $\sim15$~K, on average, consistent with the temperatures determined in other IRDCs \citep{pillai2006}. Temperatures above $T_\mathrm{K}\gtrsim20$~K are reached only in proximity of the MIR point sources and protostars (red diamond and black crosses) detected in the field. These sources could cause local heating via their radiation or outflows, whose presence was previously hinted through extended SiO emission with broad lines in the region \citep{jimenez-serra2010,liu2026-alma-sio}.

\end{appendix}

\end{document}